\documentclass[namedreferences]{solarphysics}
\usepackage[optionalrh]{spr-sola-addons} 
\usepackage{epsfig}          
\usepackage{graphicx}        
\usepackage{color}           
\usepackage{url}             
\usepackage{enumerate}
\usepackage{hyperref}
\usepackage{textcomp}

\newcommand{\etal}{\textit{et~al.}}

\newcommand{\tc}{\,^{\circ}\mathrm{C}}

\newcommand{\adv}{    {\it Adv. Space Res.}} 
 
\newcommand{\aap}{    {\it Astron. Astrophys.}}

\newcommand{\apj}{    {\it Astrophys. J.}}

\newcommand{\grl}{    {\it Geophys. Res. Lett.}}

\newcommand{\jgr}{    {\it J. Geophys. Res.}}

\newcommand{\pasp}{   {\it Pub. Astron. Soc. Pac.}}

\newcommand{\solphys}{{\it Solar Phys.}}
 
\newcommand{\ssr}{    {\it Space Sci. Rev.}} 

\newcommand{\ion}[2]{#1\,\textsc{#2}}

\newcommand{\eg}{\textit{e.g.}}

\usepackage{solaheader}	
\newcommand{\solareceiveddate}{19 September 2012} 
\newcommand{\solaaccepteddate}{26 March 2013}
\newcommand{\Xdoi}{10.1007/s11207-013-0290-z}
\renewcommand{\solayear}{2013}
\begin{document}

\begin{article}

\begin{opening}

\title{On-Orbit Degradation of Solar Instruments}

\author{A.~\surname{BenMoussa}$^{1,2}$\sep
        S.~\surname{Gissot}$^{2}$\sep
        U.~\surname{Sch\"uhle}$^{3}$\sep 
        G.~\surname{Del Zanna}$^{4}$\sep
        F.~\surname{Auch\`ere}$^{5}$\sep
        S.~\surname{Mekaoui}$^{6,17}$\sep
        A.R.~\surname{Jones}$^{7}$\sep
        D.~\surname{Walton}$^{8}$\sep
        C.J.~\surname{Eyles}$^{9,10}$\sep
        G.~\surname{Thuillier}$^{11}$\sep
        D.~\surname{Seaton}$^{2}$\sep
        I.E.~\surname{Dammasch}$^{2}$\sep
        G.~\surname{Cessateur}$^{12}$\sep
        M.~\surname{Meftah}$^{11}$\sep
        V.~\surname{Andretta}$^{13}$\sep 
        D.~\surname{Berghmans}$^{2}$\sep
        D.~\surname{Bewsher}$^{14}$\sep
        D.~\surname{Bols\'ee}$^{1,15}$\sep
        L.~\surname{Bradley}$^{8}$\sep
        D.S.~\surname{Brown}$^{14}$\sep
        P.C.~\surname{Chamberlin}$^{16}$\sep
        S.~\surname{Dewitte}$^{17}$\sep
        L.V.~\surname{Didkovsky}$^{18}$\sep
        M.~\surname{Dominique}$^{2}$\sep
        F.G.~\surname{Eparvier}$^{7}$\sep
        T.~\surname{Foujols}$^{11}$\sep
        D.~\surname{Gillotay}$^{15}$\sep
        B.~\surname{Giordanengo}$^{2}$\sep 
        J.P.~\surname{Halain}$^{19}$\sep
        R.A.~\surname{Hock}$^{20}$\sep
        A.~\surname{Irbah}$^{11}$\sep
        C.~\surname{Jeppesen}$^{7}$\sep 
        D.L.~\surname{Judge}$^{18}$\sep
        M.~\surname{Kretzschmar}$^{21}$\sep
        D.R.~\surname{McMullin}$^{22}$\sep
        B.~\surname{Nicula}$^{1,2}$\sep
        W.~\surname{Schmutz}$^{12}$\sep 
        G.~\surname{Ucker}$^{7}$\sep
        S.~\surname{Wieman}$^{18}$\sep
        D.~\surname{Woodraska}$^{7}$\sep
       T.N.~\surname{Woods}$^{7}$
       }
\runningauthor{A. BenMoussa \etal}
\runningtitle{On-Orbit Degradation}

 \institute{
$^{1}$ Solar Terrestrial Centre of Excellence, B-1180 Brussels, Belgium \\
email:\url{ali.benmoussa@stce.be}\\ 
$^{2}$ Royal Observatory of Belgium, B-1180 Brussels, Belgium\\
email:\url{samuel.gissot@oma.be} \\
$^{3}$ Max-Planck-Insitut f\"ur Sonnensystemforschung, 37191 Katlenburg-Lindau, Germany \\
$^{4}$ DAMTP - Centre for Mathematical Sciences, University of Cambridge, Wilberforce Road, Cambridge CB3 0WA, UK\\
$^{5}$ Institut Astrophysique Spatiale, CNRS-Universit\'{e} Paris-Sud 11, 91405 Orsay, France\\ 
$^{6}$ European Commission - Joint Research Centre, Ispra, Italy\\
$^{7}$ Laboratory for Atmospheric and Space Physics, University of Colorado, Boulder, CO, USA\\
$^{8}$ Mullard Space Science Laboratory, Holmbury St. Mary, Dorking, Surrey, RH5 6NT, UK \\ 
$^{9}$ Space Science and Technology Department, Rutherford Appleton Laboratory, Chilton, Didcot, UK \\ 
$^{10}$ Grupo de Astronomia y Ciencias del Espacio, ICMUV, Universidad de Valancia, Spain\\ 
$^{11}$ LATMOS-CNRS UMR 8190, Universit\'{e} Paris VI, Universit\'{e} de Versailles, France \\
$^{12}$ PMOD/WRC, Dorfstr 33 Davos, CH-7260, Switzerland\\
$^{13}$ INAF, Osservatorio Astronomico di Capodimonte, Salita Moiariello, 16, 80141 Naples, Italy\\ 
$^{14}$ Jeremiah Horrocks Institute, University of Central Lancashire, Preston, UK \\
$^{15}$ Belgium Institute for Space Aeronomy, B-1180 Brussels, Belgium\\
$^{16}$ Goddard Space Flight Center, Greenbelt, MD, USA\\
$^{17}$ Royal Meteorological Institute of Belgium, B-1180 Brussels, Belgium\\
$^{18}$ Space Sciences Center, University of Southern California, Los Angeles, CA, USA\\
$^{19}$ Centre Spatial de Li\`{e}ge, Universit\'{e} de Li\`{e}ge, Li\`{e}ge Science Park, 4013 Angleur, Belgium\\ 
$^{20}$ Air Force Research Laboratory, Kirtland Air Force Base, Alburqueque, NM, USA \\
$^{21}$ LPC2E/CNRS, (UMR 7328), Universit\'{e} d'Orl\'eans, 45071 Orl\'eans cedex 2, France\\
$^{22}$ Space Systems Research Corporation, 1940 Duke Street 200, Alexandria, VA 22314, USA \\
           }

\begin{abstract}
We present the lessons learned about the degradation observed in several space solar missions,
based on contributions at the Workshop about On-Orbit Degradation of Solar and Space Weather Instruments that took place at the Solar Terrestrial Centre of Excellence (Royal Observatory of Belgium) in Brussels on 3 May 2012. 
The aim of this workshop was to open discussions related to the degradation observed in Sun-observing instruments exposed to the effects of the space environment. 
This article summarizes the various lessons learned and offers recommendations to reduce or correct expected degradation with the goal of increasing the useful lifespan of future and ongoing space missions. 
\end{abstract}
\end{opening}

\section{Introduction}
     \label{S-Introduction} 

Investigation and analysis of the degradation of space instruments are crucial parts of achieving the scientific goals of all such instruments. 
Remote--sensing instrumentation exposed to the space environment usually degrades due to the harsh environment in which the instruments are expected to operate. 
Solar instruments  -- telescopes, spectrographs and radiometers -- are particularly vulnerable because their optical elements are exposed to unshielded solar radiation. 
For example, such instruments have historically suffered substantial degradation due to a combination of solar irradiation and instrumental contamination that can cause polymerization of organic material and, subsequently, irreversible deposition of this material on the instruments' optical surfaces. 

Different methods and approaches have been used to assess and monitor the evolution of these instruments' degradation. 
In order to reach a better understanding of how to both monitor and study this degradation,
 the Solar Terrestrial Centre of Excellence (STCE) at the Royal Observatory of Belgium organized a workshop on this subject on 3 May 2012, in Brussels, Belgium.
Representatives from several active space-based solar instruments (see Table \ref{T-instruments}) contributed to this workshop.

In this article we present analyses of these instruments' degradation (or non-degradation), the causes of degradation when they could be identified, the consequences of degradation, and methods by which the impact of degradation can be mitigated.
We also provide a summary of the lessons learned and recommendations for best practices with the hope that this information will help scientists and engineers to prevent -- or cope with -- degradation of active and future space-based solar instruments.

\begin{table}
\caption{List of the solar space instruments used. The acronyms used in this table are defined in their corresponding section.}
\label{T-instruments}
\begin{tabular}{llll}     
  \hline                   
Mission/ & Telescope (T)& Spectral range& Mission length\\
	instrument & Spectrometer (S)&[nm] &( ... -- present)\\
     & Radiometer (R)& &\\
  \hline
SOHO/SUMER & T-S & 66\,--\,161 & Dec. 1995\\
SOHO/CDS & S & 15\,--\,79 & Dec. 1995\\
SOHO/EIT & T & 17\,--\,30 & Dec. 1995\\
SOHO/CELIAS-SEM & R-S & 0.1\,--\,50 & Dec. 1995\\
SOHO/DIARAD & R & Total Solar Irradiance  & Dec. 1995\\
HINODE/EIS & T-S  & 17\,--\,29 & Sep 2006\\
STEREO/HI1A HI1B & T & 630\,--\,730 & Oct 2006\\
ISS/SOLSPEC& R-S & 165\,--\,3080 & Feb 2008\\
PROBA2/SWAP & T  & 17.4 & Nov 2009\\
PROBA2/LYRA & R & 0.1\,--\,70 121.6 190\,--\,222 & Nov 2009\\
SDO/EVE: & R-S  & 0.1\,--\,105 121.6 & Feb 2010 \\
EPS, MEGSA1-A2-B-P &   &  &  \\ 
Picard/PREMOS & R & 210 215 266 535 607 782 & June 2010\\ 
Picard/SODISM & T & 215 393 535 607 782 & June 2010\\ 
\hline
\end{tabular}
\end{table}


\section{Solar Instruments onboard SOHO}
\label{S-soho}    

The {\it Solar and Heliospheric Observatory} (SOHO: \opencite{Fleck1995}) is a successful solar mission that includes -- among other instruments -- radiometers,
spectrometers and an extreme-ultraviolet (EUV) imager, and has operated for more than 16 years at the Lagrangian L$_1$ point. 
SOHO was launched in December 1995 and began routine operations in January 1996.
At the beginning of the development phase of SOHO, the degradation processes were analyzed and appropriate procedures and design concepts were developed to eliminate them. 
During the design phase of SOHO's instruments and spacecraft, 
a meticulous cleanliness program was implemented to control molecular and particle contamination \cite{Pauluhn2002}. 
A substantial part of the success of SOHO is due to the thoughtful design of the spacecraft, payload module, and instruments and a strict material selection process. 

\subsection{Cleanliness and Calibration Stability of the SUMER Spectrograph}
\label{Cleanliness} 
Among the suite of remote-sensing instruments onboard SOHO are three spectrographs operating in the vacuum and extreme ultraviolet (VUV--EUV) range.
The {\it Solar Ultraviolet Measurements of Emitted Radiation} (SUMER) instrument is a telescope/spectrometer. 
The detailed design has been described by \inlinecite{Wilhelm1995}. In brief, it consists of a single telescope mirror and a spectrograph. 
The reflective optics, the telescope mirror, the collimator and wavelength scan mirror, and the grating are made of silicon carbide. 
The spectrograph carries two detectors with an instantaneous spectral range of 4 nm in first order (the second order spectrum is superposed). 
A wavelength-scanning mechanism selects the displayed spectral band in the range from 66 up to 161 nm.

\subsubsection{Cleanliness Program}
During the design phase of SUMER, a thorough cleanliness control program was 
implemented \cite{Schuehle1993}. 
The cleanliness requirements were estimated by model calculations of contaminants on the optical system. 
Particle fallout rates in cleanrooms published at that time were used to calculate exposure time of flight hardware inside cleanroom environments.
 With the results of the studies, a cleanliness control plan was established that contains the cleanliness requirements, 
handling practices for all hardware, and the control procedures and verification of cleanliness. 
The main features of the cleanliness program, however, were cleanliness design, material selection, cleaning, and bake-out for space conditioning. 

\subsubsection{Cleanliness Design}
The SUMER structural housing is made of aluminum that could be thoroughly cleaned and hermetically sealed when it was  
assembled. 
Only an aperture-door mechanism could open the optical compartment to the environment at any time of the assembly, integration, tests, and validation phases. 
The spring-loaded door lid provided the functionality of a valve that opens during launch for depressurization. 
A transparent window inside the door lid provided heat input by the Sun to the primary mirror when the door was closed. 
In this way, the primary mirror stays the hottest element inside the instrument. 
Internally, the telescope and spectrograph form two compartments separated by walls containing the spectrograph entrance slit.  
The two compartments were connected by large venting holes, to avoid a pressure difference between them and to prevent gas flowing through the slit.
Electronic components inside the housing are minimized. 
Apart from the optics, the mechanical driving mechanisms contain ultra-high vacuum motors and position encoders, temperature sensors, and limit switches. 
To deflect the majority of the solar-wind particles, deflector plates are implemented inside the entrance baffle, far ahead of the mirror. 

\subsubsection{Material Selection}
It was established that the usual outgassing properties of known space-qualified materials -- the Total Mass Loss (TML) and the Collected Volatile Condensable Materials (CVCM) values -- 
were not adequate to determine whether outgassing of organic material were sufficiently low to be acceptable.
A contamination study was carried out to simulate the deposition of organic material and its polymerisation under vacuum
 and UV radiation.
In addition, outgassing investigations of materials and components were conducted with gas-chromatography/mass-spectrometry (GC/MS) analysis
 to determine the outgassing of organic species as a function of temperature and time. 
The species were enclosed in a glass vial that allows the specimen to be heated in an oven and purged by clean gas over extended periods of time,
 while occasionally gas samples were drawn for GC/MS analysis. 
This procedure was particularly useful as it revealed either the rejection or the acceptance of the component. 
The specimens were heated to the highest temperature compatible with their specifications and the duration was extended until acceptance was achieved,
 usually when the outgassing of organic molecules was near the detection limit. 
Since this method is more sensitive than residual-gas analysis under high vacuum, it was adopted to design the bake-out procedure in a similar way.

\subsubsection{Cleaning and Bake-out}
Generally, all hardware used for assembly was precision cleaned before entering the cleanroom facilities.
The cleaning procedures included, with only a few exceptions, an ultrasonic bath with detergent and ultra-clean water. 
Solvent-compatible items were also cleaned with isopropyl alcohol and acetone. 
A special procedure was applied to cables before production of harnesses with methyl--ethyl ketone (now replaced by a special detergent), 
to remove possible residues of silicons inside the cable insulation. 
After wet cleaning, items selected for integration were subjected to a bake-out procedure and transferred to the clean area in double bags.
The bake-out oven consists of a chamber with a controlled heating system and vacuum port for pump-out. 
All items were baked at the highest temperature compatible with the material of construction. 
Sub-assemblies were either baked at component level before assembly or, if subjected to an outgassing test as described above, baked at the temperature and duration determined by the test. 
This bake-out procedure replaced the usual space conditioning under high vacuum with heating and collecting contaminants at cold plates.\\
To guarantee the cleanliness of the six motor-driven mechanisms on SUMER, only dry lubrication of bearings with sputtered molybdenum disulphide (MoS$_2$) was considered acceptable.
The motor coils were baked in an oven under clean gas purging (N$_2$ grade 5.0) at $200\tc$ for 48 hours before assembly of the component.

\subsubsection{Ground Calibration}
The responsivity of SUMER was characterized in the laboratory with a transfer source standard calibrated by the Physikalisch-Technische Bundesanstalt (PTB) 
at the {\it Berlin Electron Storage Ring for Synchrotron Radiation} (BESSY II: \opencite{hollandt1996}).
 The transfer source is based on a hollow cathode (HC) discharge source, operated with inert gases to deliver a number of spectral lines inside the SUMER spectral range.
 However, the range could not be covered continuously and this left some gaps in the calievebrated wavelength range \cite{Schuehle1994}.
Recently, the PTB has opened an electron storage ring, the {\it Metrology Light Source} (MLS: \opencite{klein2008} and \opencite{Gottwald2010}),
which operates with a continuous spectrum and with capabilities to calibrate space instruments.

\subsubsection{Onboard Calibration Tracking}

\begin{figure}    
   \centerline{\includegraphics[width=1\textwidth,clip=]{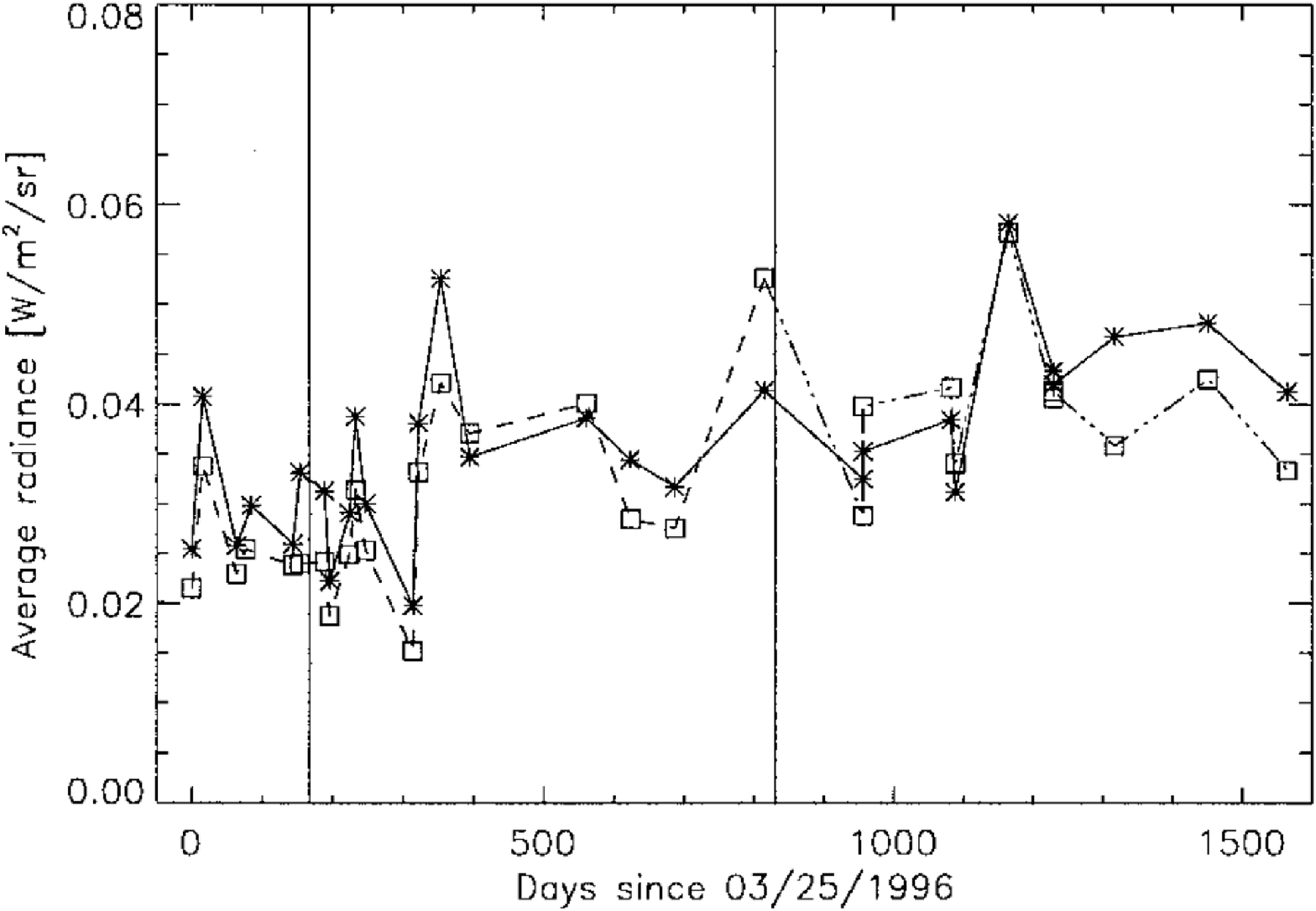}}
              \caption{Common observations of the \ion{Mg}{x} spectral line emission at 62.4\,nm in quiet-Sun areas by CDS (asterisks) and SUMER (squares) during the first years of SOHO operations \protect\cite{Pauluhn2001}.}
   \label{fig:sumer1}
   \end{figure}

It is very important to track any degradation during the time of the mission. 
For SUMER it was possible to observe repeatedly UV-bright stars that come into the field of view (FOV) every year \cite{Lemaire2002}.
Another way of tracking is the observation of the radiance of quiet-Sun areas not affected by active regions with large variability. 
The radiance of these quiet-Sun areas have been shown to vary only slowly over time periods of a solar-activity cycle \cite{Schuehle1998}. 
By observing the same objects simultaneously in common wavelength ranges, this method can be used for inter-calibration between instruments. 
This has been done successfully over several years between UV instruments on SOHO \cite{Pauluhn2002}. 
The {\it Coronal Diagnostic Spectrometer} (CDS) and SUMER spectrometers have made such common observations from the start of their operational phases. 
Figure \ref{fig:sumer1} (taken from \opencite{Pauluhn2001}) depicts clearly that, by common observation of the quiet Sun, 
the degradation of the two instruments could be well accounted for, such that the remaining variation of the signal is not a systematic error of the instrumental throughput.
This common intercalibration procedure, however, does not take into account longer-term effects on the CDS responsivity, discussed in the following section.

\subsection{The Calibration of CDS}

CDS is composed of a {\it normal 
incidence spectrometer} (NIS) and a {\it grazing incidence spectrometer} (GIS : \opencite{harrison95}).
The two instruments share  a  Wolter--Schwarzschild type II  grazing-incidence 
telescope, a scan mirror, and a set of different slits. There is no entrance 
filter. GIS uses a spherical grating that disperses the incident 
light into four spiral anode (SPAN) microchannel plate (MCP) detectors.\\
The NIS is composed of two stigmatic toroidal gratings that 
disperse the  radiation into two wavebands (NIS 1: 30.8\,--\,37.9~nm and NIS 
2: 51.3\,--\,63.3 nm).
The NIS detector comprises an MCP Philips model G12-33 
with pores of 12\,\textmu m diameter. 
The EUV photons are converted into electrons via the photoelectric effect on 
the front face of the MCP, and then amplified at about 756 V.
The electron cloud is proximity focused onto a P-20 phosphor coated on a fiber-optic output window. The visible phosphorescence is focused via a lens onto a 
Tektronix $1024 \times 1024$ charge-coupled device (CCD) with square pixels of 21\,\textmu m.
The CCD is running cold, at a nominal temperature of -70$\tc$.

\subsubsection{Calibration}
About two years before launch, the CDS instrument was calibrated end-to-end at the Rutherford Appleton Laboratory (RAL)
against a ``transfer'' source that was absolutely calibrated using synchrotron emission. 
Details can be found  in \inlinecite{lang_etal_00}. 
Immediately after launch, it became obvious that large departures (factors of two to three) from the pre-launch calibration were present.
On 15 May 1997, a  National Aeronautics and Space Administration (NASA)/Laboratory for Atmospheric and Space Physics (LASP) rocket carried an {\it EUV Grating Spectrograph} (EGS) that had been calibrated against synchrotron emission.
On the same day, NIS measurements were performed and compared to the EGS ones \cite{brekke_etal:00}, providing one key element in the  long history of the in-flight calibration;
see  \inlinecite{delzanna_etal:10_cdscal} for a summary.
Further information on the NIS 1 was obtained with the Solar EUV Rocket Telescope and Spectrograph (SERTS)-97 rocket flight \cite{delzanna01_cdscal}. 
The only in-flight radiometric calibration of all the nine CDS channels (three second-order) was obtained by \inlinecite{delzanna01_cdscal} with the line ratio technique.
The various NIS first-order calibrations were consistent, within 30\,\% to 50\,\%, 
with the SUMER calibration, as discussed during two ISSI workshops \cite{lang_etal:02}, and as summarized in the previous section.
Further EGS rocket flights were flown in 2002, 2003, and 2004 but were not useful for the CDS calibration.
An update to the CDS radiometric calibration was instead made possible by two {\it Extreme-Ultraviolet Normal-Incidence Spectrograph} (EUNIS) rocket flights which took place in 2006 and 2007 \cite{wang_etal:11}.

\subsubsection{Detector Degradation and Long-Term Aging}
MCPs are known to suffer a drop in  gain owing to the exposure to solar radiation. 
For the NIS, this  results in a depression at the core of the lines caused by exposures with the 2 arcsec or 4 arcsec slits (the so-called ``burn-in'' of the lines).
This effect can be  corrected by looking at the burn-in in 90 arcsec slit exposures of the quiet-Sun \cite{thompson:00}.  

\begin{figure}
 \centering
   \includegraphics[width=1\textwidth]{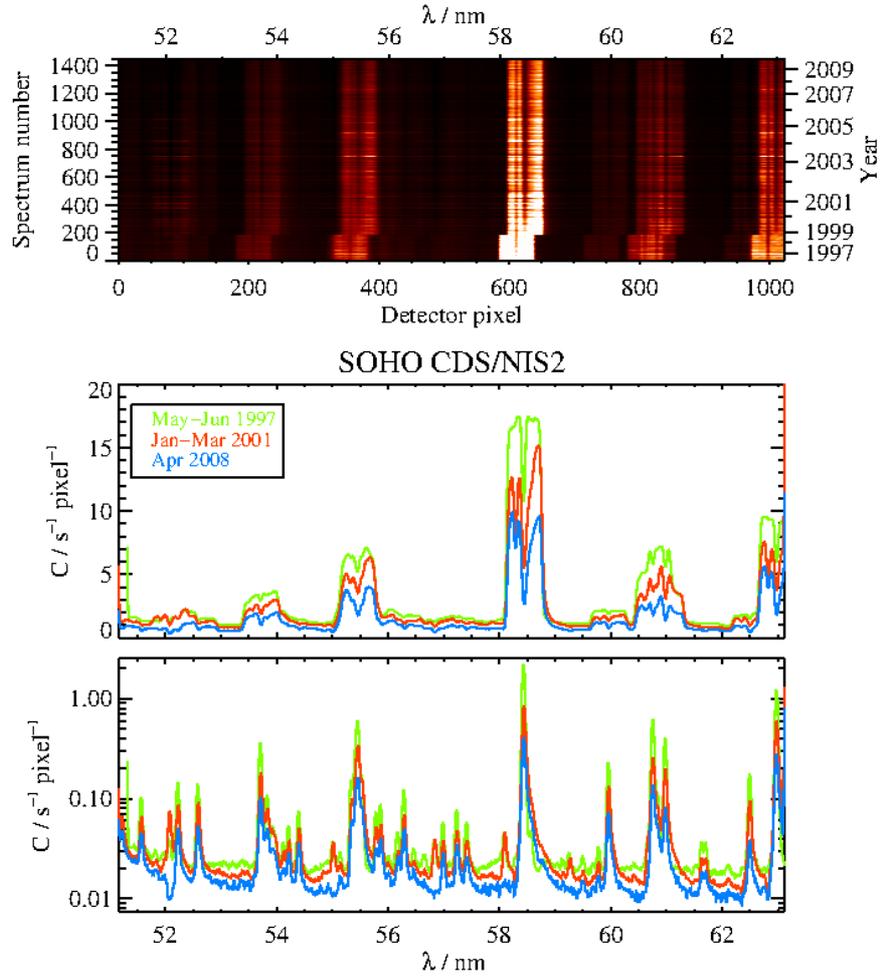}
  \caption{%
    Top: the 13 years of CDS 90 arcsec NIS~2 data over the
    quiet-Sun; the last spectrum shown here was taken on 7 April 2009.
    Middle: A sample of averaged 90 arcsec spectra 
    at three different epochs.
    Bottom: The corresponding 2 arcsec spectra.  Notice the overall
    decrease in the count rates [$C$] in both 2 arcsec, and 90 arcsec spectra, and
    the marked decrease of count rates corresponding to the core of strong
    lines such as, for example, the ion He I 58.4 nm line. Figure adapted from \protect\opencite{delzanna_etal:10_cdscal}.
      }
  \label{fig:nis_s6_image}
\end{figure}

Figure \ref{fig:nis_s6_image} shows such a burn-in effect in a 13-year long series of near simultaneous spectra taken with the narrow 2 arcsec and the wide 90 arcsec slits.

It was thought that exposing with the 90 arcsec slit would
significantly reduce (by more than a factor of three
over 13 years) the responsivity at the wavelengths where the stronger lines
in the spectra are present (\opencite{thompson:00}; 2006).
However, as shown in a series of articles (see references in 
\opencite{delzanna_etal:10_cdscal}), this assumption turned out to be incorrect, 
in that an overall decrease across all wavelengths of about a factor of two in 13 
years was measured. Some wavelength-dependent effects turned out to be minor.

\begin{figure}
  \centering
  \includegraphics[angle=90, width=1\textwidth, clip=false]{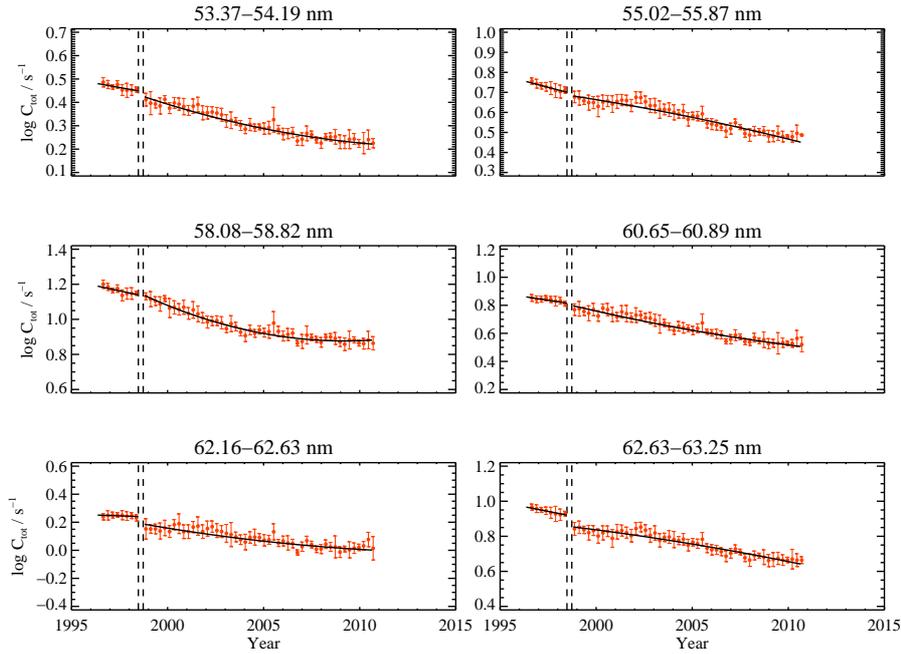}
   \caption{Time-dependence of average radiances in various wavelengths from CDS
     90 arcsec quiet-Sun observations in the NIS-2 channel.
     Line radiances from individual spectra were grouped in bins of 90 days, 
     obtaining an estimate of the mean and standard deviation in each time bin (points and bars).  Black curves represent fits to these data. The two vertical
dashed lines represent the times of loss (25 June 1998) and recovery
(25 September 1998) of contact with SOHO, respectively.}
         \label{fig:nis2_nimcp_s6}
 \end{figure}

The overall decrease and its magnitude over more than 13 years of
monitoring, are readily apparent in the spectra shown in Figure~\ref{fig:nis_s6_image}.
Note in the top panel of Figure~\ref{fig:nis_s6_image} the discontinuity due to the temporary loss of contact with SOHO that occurred in June through September 1998.
The resulting exposure to rather extreme and uncontrolled environmental conditions during that time interval caused significant changes in the characteristics of CDS,
 as of most other SOHO instruments. 
Figure~\ref{fig:nis2_nimcp_s6} illustrates this effect on the NIS sensitivities and shows the measured aging in various wavelengths of the NIS~2 channel from 1996 to September 2010.  
The coefficients of the fits to these data have been adopted as the
default long-term radiometric correction in the latest version of the CDS
analysis software. Figure~\ref{fig:nis2_nimcp_s6} is similar to Figure 4 of \inlinecite{delzanna_etal:10_cdscal},
but shows data processed after the publication of that article, until September 2010.
The curves shown in Figure~\ref{fig:nis2_nimcp_s6} therefore reflect the most up-to-date estimates of the long-term variation of the CDS/NIS
 sensitivity, as currently included in the CDS analysis software.
  
The characterisation of the long-term aging was found by 
\inlinecite{delzanna_etal:10_cdscal} assuming that the quiet-Sun radiances in 
low-temperature lines are constant over time.
Some support for this assumption comes from ground-based measurements of 
equivalent widths of photospheric and chromospheric lines (\eg{} \ion{Ca}{ii}) 
over the quiet Sun (often Sun-center), which have provided firm  evidence 
that the basal photospheric-chromospheric emission has not changed over the 
past three solar cycles (\opencite{livingston_etal:07}; \citeyear{livingston_etal:10}).

The validity of the assumption has been confirmed by the overall agreement 
found between the CDS and the Solar Dynamics Observatory/EUV Variability Experiment (SDO/EVE) 2008 prototype irradiances 
\cite{delzanna_etal:10_cdscal} and by a direct CDS-EUNIS rocket flight 
comparison of radiances \cite{wang_etal:11}.
Note that this assumption does not necessarily contradicts reports of an intrinsic variation of the solar source (see Section 2.1.6), 
whose magnitude typically is much smaller than the extent of the long-term instrumental degradation,
and might be affected by the increase in solar activity in the 1996\,--\,2001 period considered by those studies (\eg{} by \opencite{Pauluhn2003}).

\inlinecite{delzanna_andretta:11} proposed a new calibration for the \ion{He}{ii} line, providing CDS irradiances in excellent 
agreement with those measured by the SDO/EVE prototype and by EUNIS \cite{wang_etal:11}.
As described by \inlinecite{Kuin2007}, there were gain-depression effects 
that lowered the spectral resolution in the strongest GIS lines. However, 
overall no signs of a decrease in responsivity were observed over a time period 
of ten years.
This suggests that the grazing-incidence optics (telescope, common
to the NIS) and the GIS grating have not suffered any contamination.
It is therefore likely that the slow decrease in responsivity experienced by 
NIS is due to an overall decrease in the reflectivity of the normal-incidence 
gratings, or to a lower sensitivity of the detector, either in the CCD or in the phosphorous 
coating on the anode in front of the MCP.

\subsection{In-Flight Evolution of EIT}
The {\it Extreme Ultraviolet Imager} (EIT) onboard SOHO is a Ritchey--Chretien telescope observing the Sun in four passbands of the EUV spectrum: 
17.1 nm (\ion{Fe}{ix/x}), 19.5 nm (\ion{Fe}{xii}), 28.4 nm (\ion{Fe}{xv}) and 30.4 nm (\ion{He}{ii}, \ion{Si}{xi}). 
Four different multilayer coatings on the primary and secondary mirrors are used to select the passbands. 
A sector wheel at the front of the instrument is used to select one of the four quadrants. 
Thin film aluminum (Al) filters at the entrance of the instrument suppress the incoming visible and infrared (IR) radiation. 
Additional filters at the focal plane and on a filter wheel provide redundancy. 
The $1024 \times 1024$ CCD detector is passively cooled to about -70\,\textcelsius. A shutter is used to time the exposures. 
A detailed description of the instrument can be found in \inlinecite{Delaboudiniere1995}.
Since its first light in January 1996, EIT has provided revolutionary views of the EUV Sun  \cite{Moses1997}. 
Its observations have been affected by serious degradation issues, but the degradation process could be understood and corrected for.
A detailed analysis of the in-flight performances of EIT is given by \inlinecite{Defise1999}.

\subsubsection{Detector Degradation and In-Flight Correction} 
\begin{figure}    
   \centerline{\includegraphics[width=1\textwidth,clip=]{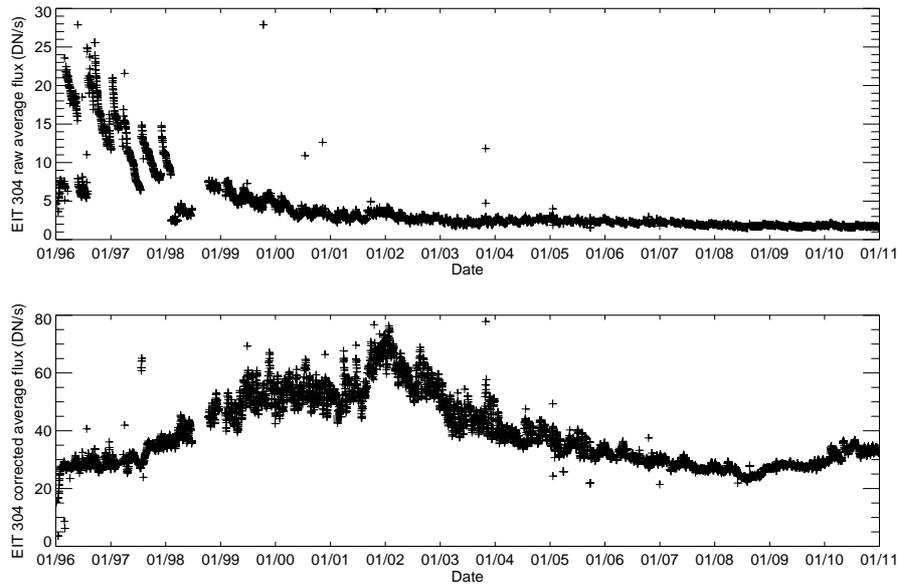}}
              \caption{Average flux in the \ion{He}{ii} 30.4 nm band of EIT as a function of time before (top) and after (bottom) correction of the degradation.}
   \label{fig:eit2}
   \end{figure}

The total flux in EIT images rapidly showed strong variations that were obviously uncorrelated from the solar activity, as shown in Figure \ref{fig:eit2}.

 We see an initial rise of the signal followed by a steep decrease and periodic discontinuities.
 The initial rise is attributed to rapid outgassing of the instrument following the opening of its sealed door. 
There are two causes for the subsequent decrease: absorption by a contaminant on the detector surface and radiation-induced degradation of its charge collection efficiency (CCE). 
For operational reasons, the partial pressure in the vacuum vessel was not low enough at launch so that immediately after launch, water condensed on the rapidly cooling detector.
Modeling showed that the observed absorption could be explained by a thin layer of water, which is consistent with insufficient pumping at launch \cite{Defise1999}. 
Other contaminating compounds could, of course, also be present,
but we have no way of identifying them in-flight, while water is known to be present in the residual atmosphere present in the vacuum vessel at launch.
For stray-light protection, the CCD is isolated from the rest of the telescope by a flange and the only outgassing path to space is two labyrinths of low conductance.

\begin{figure}  
 \centering {\includegraphics[width=1\textwidth]{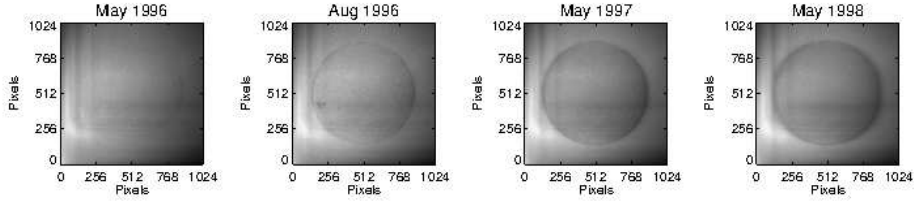}}
 \caption{Evolution of EIT's calibration lamp images during the first two years of the SOHO mission (from \protect\opencite{Defise1999}).}
\label{fig:eit3}
\end{figure}

The detector was regularly baked out to evaporate the water but because of the low conductance to the outside, 
most of the water simply condensed on the walls of the back end and went back to the CCD when it was cooled again. 
This cycle is one cause for the oscillating response seen in Figure \ref{fig:eit2}. 
The second component of detector degradation is illustrated in Figure \ref{fig:eit3}. 
The four images show the evolution of calibration lamps (a light bulb illuminating the focal plane with visible light). 
We see the progressive imprinting of a negative average image of the EUV Sun, with the limb brightening and active-region bands clearly visible in the last image. 

This is due to loss in the CCE of the detector. 
EUV light creates positive charges at the interface between the silicon (Si) and the silicon oxide (SiO$_2$) layer that thicken the dead layer.
 As consequence, in the regions that are on average the most illuminated, the CCE is decreased.
 By baking out the CCD regularly to about 20\,\textcelsius, the dead layer is thinned and homogenized, thus restoring part of the original sensitivity \cite{Defise1997}. 
These detector degradation effects have been understood and empirically modeled using calibration lamp images. 
However the images of Figure \ref{fig:eit3} cannot be used directly to correct the data,
firstly because they are white-light images, and secondly because the light source does not illuminate the detector uniformly. 
What has to be used is the ratio between a calibration-lamp image and an initial-lamp image taken before the first light with a pristine detector.
 This gives a visible-light flat field. This can then be converted into an EUV flat field if one knows the relationship between visible and EUV degradation. 
To derive this relationship we used several off-point maneuvers of the SOHO spacecraft to derive the EUV flat field.

\begin{figure}    
   \centerline{
               {\includegraphics[width=1\textwidth,clip=]{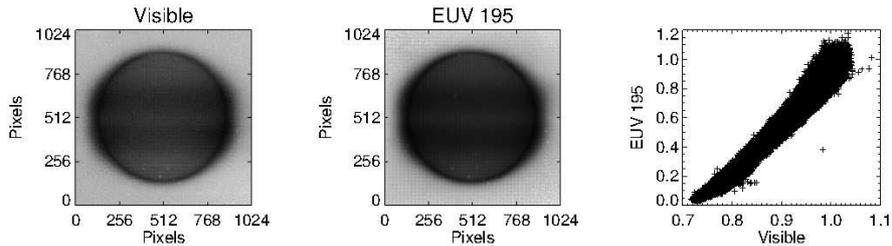}}
              }     
\caption{Left: white-light flat field for 8 February 2001. Middle: corresponding EUV flat field deduced from an offpoint of SOHO. Right: scatter plot of EUV {\it vs.} visible.}
   \label{fig:eit4}
   \end{figure}

\begin{figure}    
   \centerline{
               \includegraphics[width=0.4\textwidth,clip=]{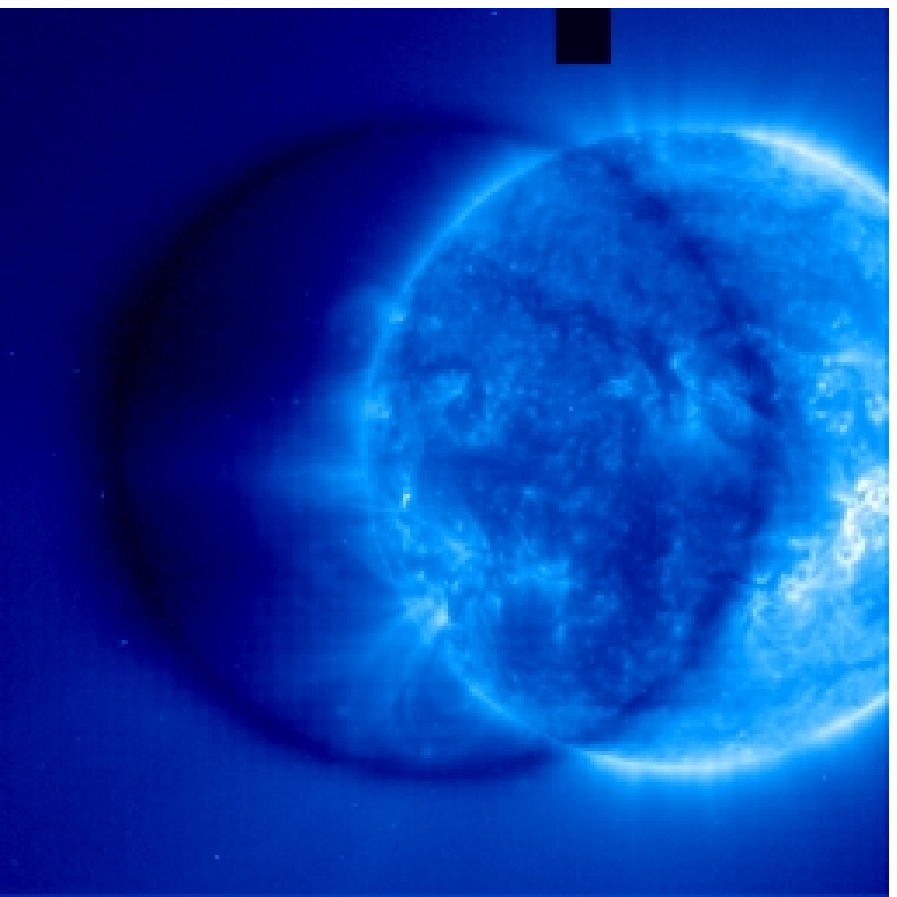}
               \includegraphics[width=0.4\textwidth,clip=]{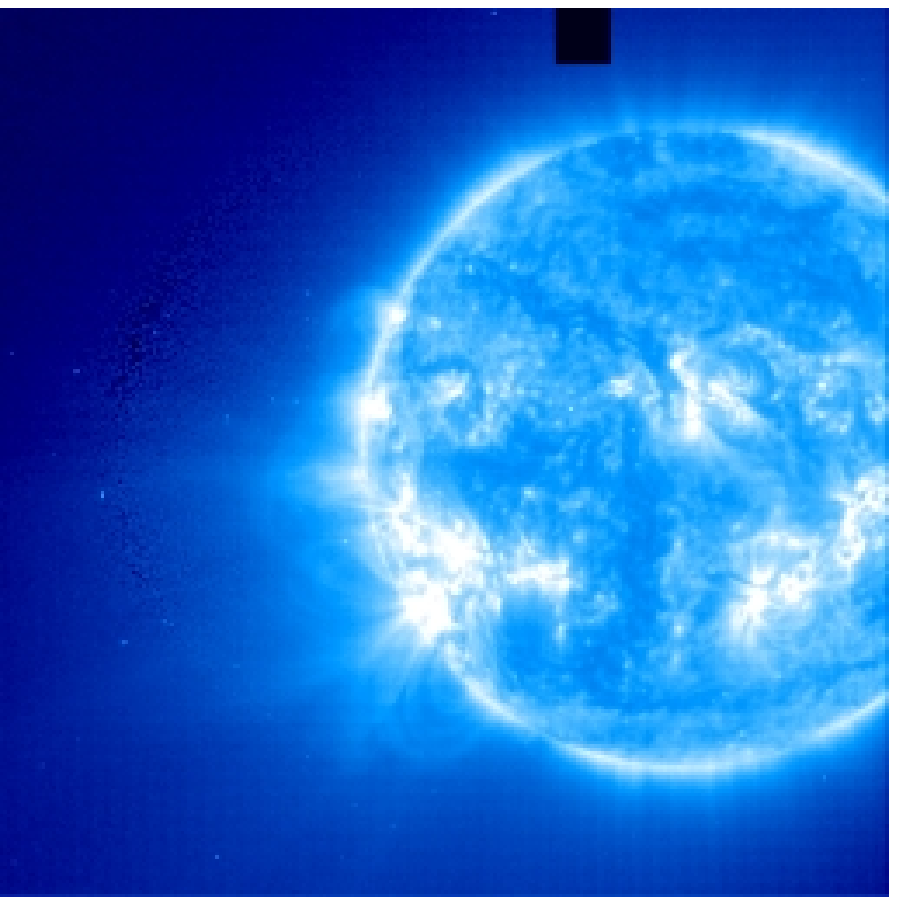}}   
   \caption{EIT 171 images taken during an off-point manoeuvre before and after correction.}
   \label{fig:eit5}
   \end{figure}

The set of displaced images taken during the maneuvers was processed using the algorithm of \inlinecite{Kuhn1991} to separate the solar image from the flat field features. 
Figure \ref{fig:eit4} shows the white-light flat field derived from calibration lamps for the date of the 8 February 2001 off-point (left),
the EUV flat field (center), and the relationship between the two (right). 
A fit to this relationship (see Figure 8.7 of \opencite{Clette2002})
is used to convert white-light calibration-lamp images taken regularly (about every two weeks) into EUV flat fields used to correct the images (see Figure \ref{fig:eit5}).
Calibration lamps have proven to be essential to the calibration of the EIT instrument. 
The key to their successful use is the acquisition before the first light of a good reference image.

Applying this to all images, we obtain a corrected EUV times series. 
The instrumental effects are taken out, revealing the solar variability.
However, the corresponding time series of integrated fluxes still exhibit semi-periodic fluctuations,
which indicates that not all of the degradation is accounted for by this procedure. 
Indeed, since the contaminant (be it only water or a mix of several compounds) is very thin, 
it is practically transparent to visible light, and therefore not revealed by the calibration lamp images. 
The onboard flat fielding thus corrects only for the CCE degradation, which is however the dominant effect in the response of EIT.
 
By comparing EIT and the {\it Solar EUV Monitor} (SEM) data, 
we concluded that the contaminant (probably essentially trapped on the cold detector) represents about 20\,\% of the total degradation \cite{Clette2002}. 
To take out this remaining variation, it was chosen to tie the EIT fluxes to the \ion{Mg}{ii} center to wing ratio index. 
For each period between two successive bake-outs, the CCE-corrected EIT integrated fluxes are correlated with the \ion{Mg}{ii} index and detrended using a linear fit. 
Details about the procedure are given by \inlinecite{Clette2002}. 
It is important to emphasize that this does not force the EIT fluxes to match a solar index; it only forces a linear relationship between the two, and the correction is 20\,\% at most.

\inlinecite{hock2008} used the {\it Thermosphere Ionosphere Mesosphere Energetics} (TIMED) / {\it Solar EUV Experiment} (SEE) spectral irradiance measurements instead of the \ion{Mg}{ii} index to correct the variations remaining after CCE correction. 
They argue that TIMED/SEE observations would be better suited because \ion{Mg}{ii} is not sensitive to coronal temperatures. 
However, the authors show that the \ion{Mg}{ii} corrected EIT fluxes agree with the TIMED/SEE measurements within the uncertainties of the two instruments. 
Furthermore, this index has the advantages of being available continuously for the entire SOHO lifetime and of being determined independently from several sources.

\subsubsection{Filter Degradation}
\begin{figure}    
   \centerline{
               \includegraphics[width=0.5\textwidth,clip=]{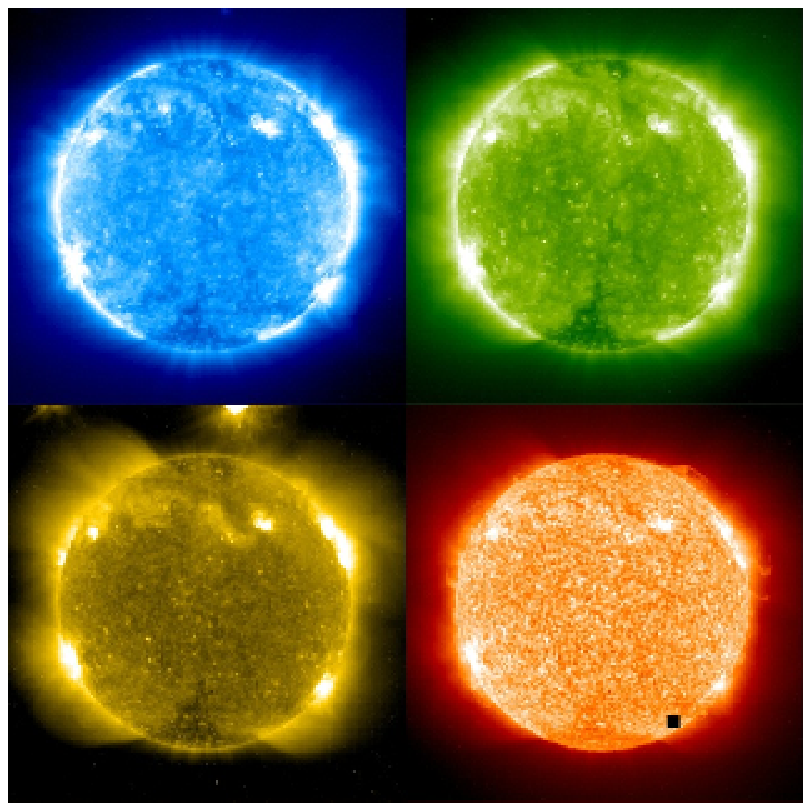}
               \includegraphics[width=0.5\textwidth,clip=]{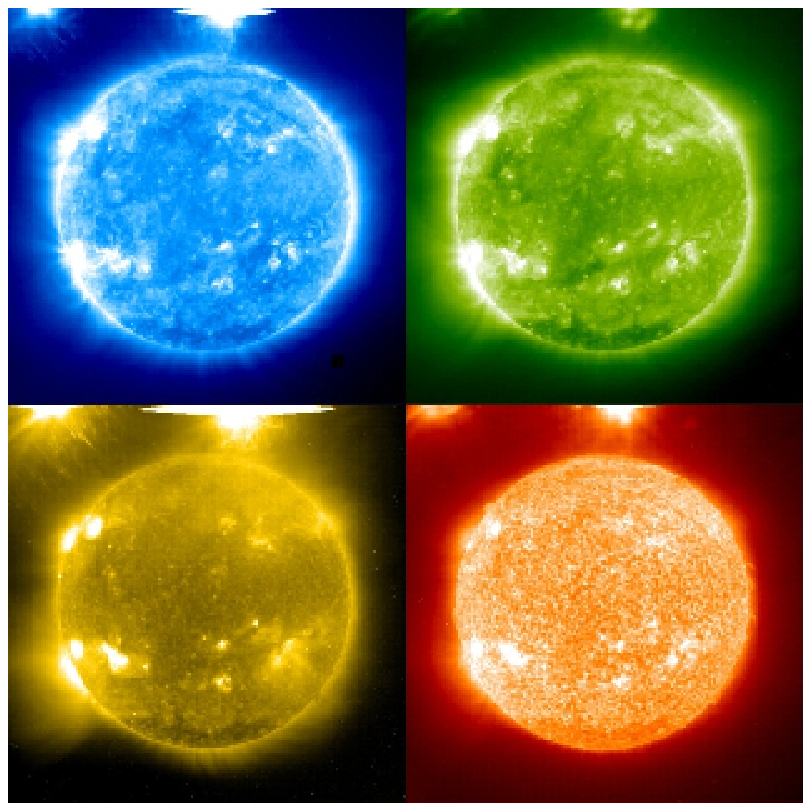}}     
\caption{Visible light leaks in EIT images. Left: after launch. Right: after the micrometeorite event of February 1998.}
   \label{fig:eit7}
   \end{figure}

The Al filters at the front of the instrument and at the focal plane did not survive the launch even though the telescope was launched under vacuum to avoid acoustic vibrations.
Tears developed, which produced light leaks that were fortunately localized to the edges of the detector so that they did not significantly affect the image quality (Figure \ref{fig:eit7}, left).
However, in February 1998, after two years in orbit, the light leaks suddenly (from one image to the next) became much larger and the images were swamped by white light (Figure \ref{fig:eit7}, right). 
The most probable explanation for this is a micrometeorite hit that produced a large pinhole in the front filters. 
 Large amounts of white light could thus reach the focal plane, and reach the detector through the pinholes that formed in the back filter during launch. 
The solution to this problem was to insert in the beam one of the extra Al filters held by the filter wheel.
Without this mechanism, EIT images would be almost unusable after 1998. 
This lesson should be remembered for future long-duration missions. 
Not only can the launch be harmful to these filters, but sudden degradation can occur at any time, especially if the spacecraft encounters harsh environments. In this case, the redundancy provided by a mechanism is essential. 

\subsubsection{Conclusion}
Sixteen years of EIT observations have given us many examples of the issues that can be encountered during the lifetime of an EUV telescope. 
EIT shows no significant changes in its spectral selectivity, but variability of the detector sensitivity was a major hurdle. 
Part of this problem (the contamination by water) could have been avoided if the pressure in the vacuum vessel had been maintained low enough up to the launch.
This illustrates the importance of maintaining cleanliness during the whole lifetime of the instrument.
The contaminants were also trapped in the back end of the telescope due to insufficient conductivity to the telescope section and to space.
This issue was solved on Solar TErrestrial RElations Observatory/Extreme UltraViolet Imager (STEREO/EUVI) by adding vents close to the detector and this practice should be maintained for future telescopes.
Comparing with other instruments, like CDS or SEM onboard SOHO, we concluded that most of the variability of the instrument response (70\,\%) is explained by the degradation of the detector.
This behavior is different from that of SEM, even though the two instruments observe in comparable wavelength bands.
The degradation observed in SEM is explained in terms of a carbon deposit on the front filter,
while on EIT there is no evidence of variation in the EUV response of either the filters or the multilayer coatings.
This difference may find its source in a combination of factors such as different designs, materials, contamination control plans, locations on the spacecraft, {\it etc}.

\subsection{SOHO/CELIAS-SEM}
\label{SEM}  
The Solar and Heliospheric Observatory / Charge, Element, and Isotope Analysis System - 
Solar EUV Monitor (SOHO/CELIAS-SEM: \opencite{sem}) is a simple transmission grating spectrophotometer using an entrance Al filter to restrict the bandpass incident on the grating, and defining the bandpass of the zero-order signal. 
Detectors in the first-order are positioned to measure the 26\,--\,34 nm region of the solar spectrum, including the \ion{He}{ii} emission at 30.4 nm.

\begin{figure}[ht]
\centering
\includegraphics[width=1\textwidth, clip=false]{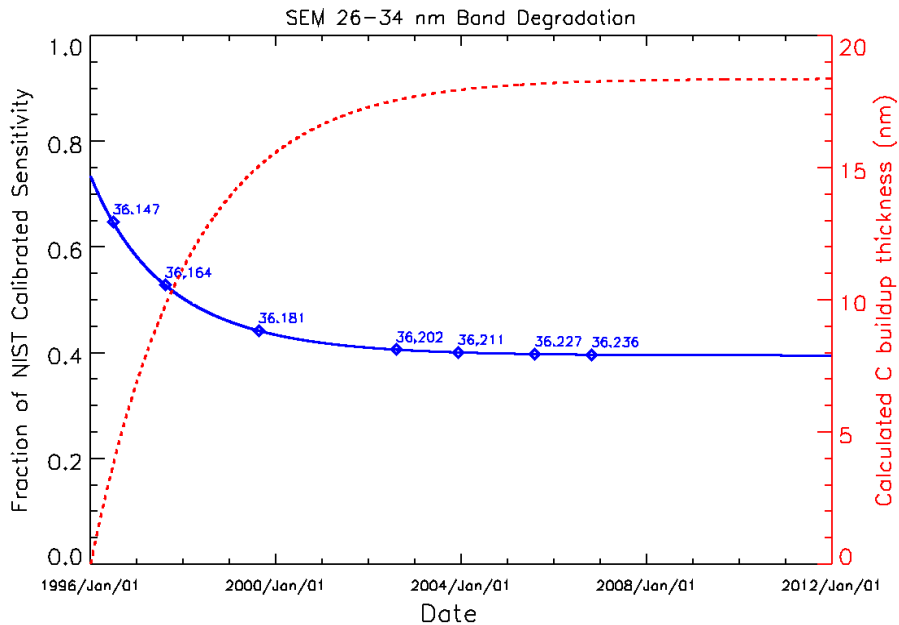}
 \caption{Degradation of the SOHO/CELIAS-SEM 26\,--\,34~nm band (blue, solid line) as measured by sounding-rocket underflights of a NIST calibrated copy of the SEM (diamonds). 
The degradation is modeled as the build-up of a layer of carbon (red, dotted line). This is an updated version of the degradation trending first presented by \protect\inlinecite{McMullin2002}.}
\label{fig:sem} 
\end{figure} 

The SEM showed steady degradation of the first-order signal over the first seven years of operation, 
and after that the degradation has remained almost constant as shown in Figure~\ref{fig:sem}. 
The degradation has been tracked by a series of sounding-rocket underflights 
with a copy of the SEM instrument that is calibrated at the National Institute of Standards and Technology (NIST) with the Synchrotron Ultraviolet Radiation Facility (SURF) before and after flight. 
This way the calibration of the sounding rocket can be applied to the on-orbit SEM, and the degradation measured.

It is postulated that the degradation seen by SEM is due to the build-up of a contamination layer on the front filter of the instrument. 
As no spectral information is available, it has been assumed that the major element causing contamination is carbon. 
Hydrocarbons from spacecraft outgassing, fuel return {\it etc.} can hit the front filter; a certain proportion will ``stick'' and can become polymerized by the solar UV radiation. 
As this layer grows, the EUV signal is more strongly attenuated.
Figure~\ref{fig:sem} shows the degradation as measured by the sounding-rocket underflights and the modeled thickness of carbon required to cause this level of attenuation in the 26\,--\,34 nm band.

\subsection{Long-Term Exposure Correction of VIRGO/DIARAD}

The {\it Differential Absolute Radiometer} (DIARAD:~\opencite{Dewitte2004}), is one of the radiometers of the  {\it Variability of solar IRradiance and Gravity Oscillations} (VIRGO: \opencite{Frohlich1995}) package on SOHO  developed at the Royal Meteorological Institute of Belgium.
 VIRGO/DIARAD has measured the Total Solar Irradiance (TSI) since 1996.
The instrument is a dual-channel, side-by-side, self-calibrating absolute radiometer. 
Each channel is composed of a detector assembly, a cylindrical black-painted cavity with its associated precision aperture, a baffling system, a limiting aperture, and a shutter.
Both cavities are mounted on a common heat sink. 
The DIARAD working principle is based on the compensation of heat in one channel's cavity while its shutter opens and closes every three minutes \cite{Dewitte2004}. 
When the shutter of the measuring channel is open, part of the solar radiation is absorbed by the cavity. Its induced heat flux is measured by the detector.
 When the shutter closes, a servo system compensates for the deficit of radiative power by dissipating an equivalent electrical power. 
When measuring with one channel, the other channel is used as a reference and its shutter is kept closed. 
For a detailed description of the instrument, see \inlinecite{Crommelynck1982}, \inlinecite{Crommelynck1999}, and \inlinecite{Mekaoui2010}.

\subsubsection{Degradation Monitoring Strategy}
The instrument being symmetric, each channel can be used as an independent measuring device with its own electrical, thermal, geometrical, and optical characteristics.
These characteristics are the main parameters in determining the absolute value of the TSI.
This last topic is still a matter of debate (\opencite{Mekaoui2010}; \opencite{Frohlich2011}; \opencite{Kopp2011a}; \opencite{Kopp2011b}; \opencite{Fehlmann2012}).
While each channel of DIARAD is electrically self-calibrating, assuring the stability of the measurements of the absolute TSI value,
 the thermo--optical properties are subject to degradation and changes due to the long-term exposure to solar radiation.

\begin{figure}[!h]
\centering
\includegraphics[width=1\textwidth, clip=false]{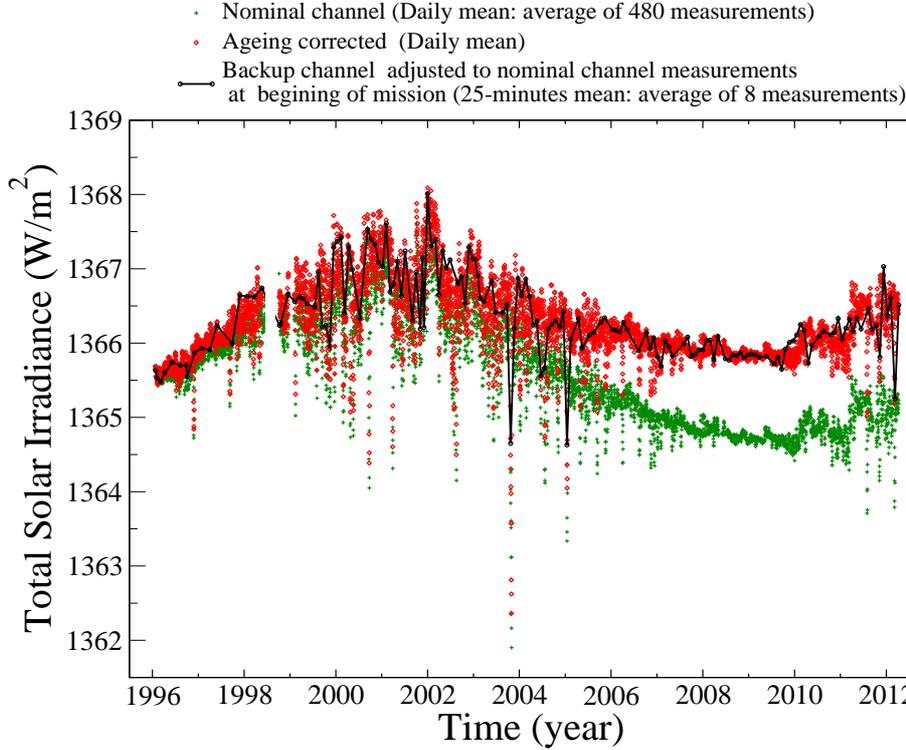}

\caption{Temporal variation of TSI measurements from VIRGO/DIARAD. Green symbols: daily mean TSI measurements from the left channel; Black curve the 25-minutes average of the TSI measurements made each month with the right channel; 
Red symbols the daily mean aging-corrected TSI measurements.}
\label{fig:both}
\end{figure}

This type of degradation is monitored and corrected by ground processing. To achieve it, the left-channel has measured the TSI every three minutes, since 1996. 
During the three minutes, the left shutter is open for only half of the time. As a consequence, the total induced aging effect from the start of the measurements in 1996 --
 assuming its unique dependence on the exposure time -- is caused by around 8.5 years of cumulative exposure to solar radiation.
The right channel is operated for only 45 minutes every month. It is exposed for half of the time to solar radiation. Its total exposure accounts for three days from the start of the mission. 
Figure \ref{fig:both} shows the daily mean TSI measurements from the left channel (in green) and the 30-minute average of the right-channel measurements
 (in black, the first 15 minutes are excluded due to a transient effect).
In this figure, the right-channel measurements have been adjusted to the left-channel measurements at the begining of the mission in 1996. 
In 2012, the difference between the two channels measurements is around 1.2\,W\,m$^{-2}$ due to the degradation of the left-channel.

\subsubsection{Degradation Correction Implications}
The less-exposed channel gives valuable information on the long-term TSI evolution. From Figure \ref{fig:both}, 
the right channel indicates that the difference between TSI minima in 2009 and 1996 is 0.15\textpm0.17\,W\,m$^{-2}$. 
This suggests that no significant increase is measured between Solar Cycle 23 minima.
Alternatively, these measurements can be used to correct the nominal (left-channel) for its aging. 

\begin{figure}
\centering
\includegraphics[width=1\textwidth, clip=false]{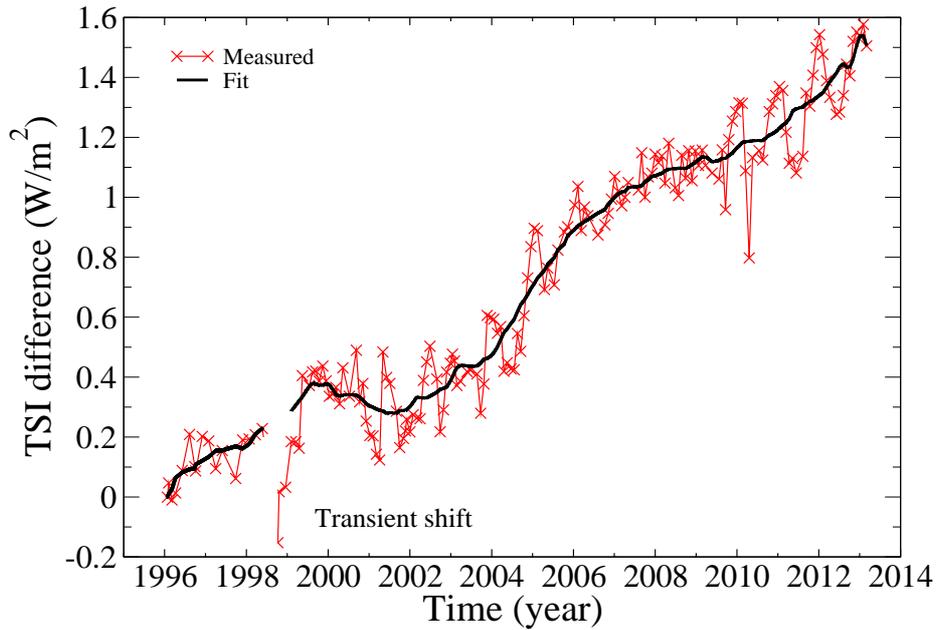}
\vspace{0.5cm}
\caption{DIARAD right minus DIARAD left TSI measurements. The offset is removed at the beginning of the mission. 
The black curve is the computed aging correction from a variable nine-point running mean over the right--left difference.}
\label{fig:aging}
\end{figure}

For each monthly measurement with the right channel, left-channel measurements are made before and after. 
These measurements are then interpolated and compared to the right-channel measurements (simultaneous measurements with the right and the left channel are not possible).
In order to reduce the uncertainty, it is important to compare the same means. 
Indeed, the right-channel measurement is a 25-minute average and so should be the left-channel measurements before the interpolation.
Figure \ref{fig:aging} shows the difference between the right-channel measurements and the interpolated left-channel measurements. A fitted curve is used to smooth the difference. 
The values of this curve are then added to the left channel measurements to take into account its degradation.
 This exposure-dependent aging is the only correction applied by the team of the Royal Meteorological Institute who developed the instrument. 
Additional corrections are applied by the VIRGO PI-team. 
These non-exposure dependent corrections are based on the comparison with other radiometers \cite{Frohlich2003}. 
These results are still a matter of debate and have yet to be reproduced on the ground.

\section{Degradation of the Hinode/EIS Detectors after Five Years in Orbit}

The {\it Hinode} satellite was launched in September 2006 and is still operational. 
{\it Hinode} is a Japanese satellite with payloads funded by JAXA--ISAS, NASA, ESA and UKSA (previously STFC). 
{\it Hinode} is in a Sun-synchronous low-Earth orbit (altitude $\approx$ 600 km), which allows for continuous observing of the Sun. 
There are three solar telescopes on-board {\it Hinode}; the {\it Solar Optical Telescope} (SOT: \opencite{Tsuneta2008}), the {\it X-Ray Telescope} (XRT: \opencite{Golub2007}), and the {\it Extreme-UV Imaging Spectrometer} (EIS: \opencite{Culhane2007}).

The {\it Extreme-UV Imaging Spectrometer} (EIS) has a large effective area in two EUV spectral bands; 17\,--\,21 nm and 25\,--\,29 nm. 
There are two CCDs, one for each wavelength range.
The CCDs are e2v device type CCD 42-20, which have an array size of 2048$\times$1024 pixels, a pixel size of 13.5$\times$13.5\,\textmu m$^2$,
are thinned for back-illumination, and employ multi-pinned phase (MPP) technology in asymmetric inverting mode operation (AIMO) which allows for low dark-current levels without excessive cooling (see \opencite{Culhane2007} for the EIS instrument article). 

The in-orbit operating temperature range of the CCDs is $\approx-36\tc$ to $-46\tc$; the variation is due to the perihelion and aphelion of the orbit.
The assembly and pre-launch calibration of EIS were performed at RAL (UK). 
The components of EIS (entrance filters, primary mirror, slit/slot mechanism, shutter, grating and CCDs) are housed in a carbon-composite structure. 
Post-launch, a pixel shift of eight\,--\,nine pixels in the spectral direction was observed when compared with the pre-launch calibration. 
The shift was attributed to the thermal stabilisation and out-gassing of the instrument, and a correction was made in the software to accommodate the shift. 
Regular calibration studies are run weekly, monthly or quarterly depending on the type of study, which include: 
dark exposures, Light Emitting Diodes (LEDs) flat-fields, synoptic, quartz crystal microbalance (QCM), and full CCD spectral scans.

\subsection{Hot and Warm Pixels}
\begin{figure}
   \centering
   \includegraphics[width=1\textwidth, clip=false]{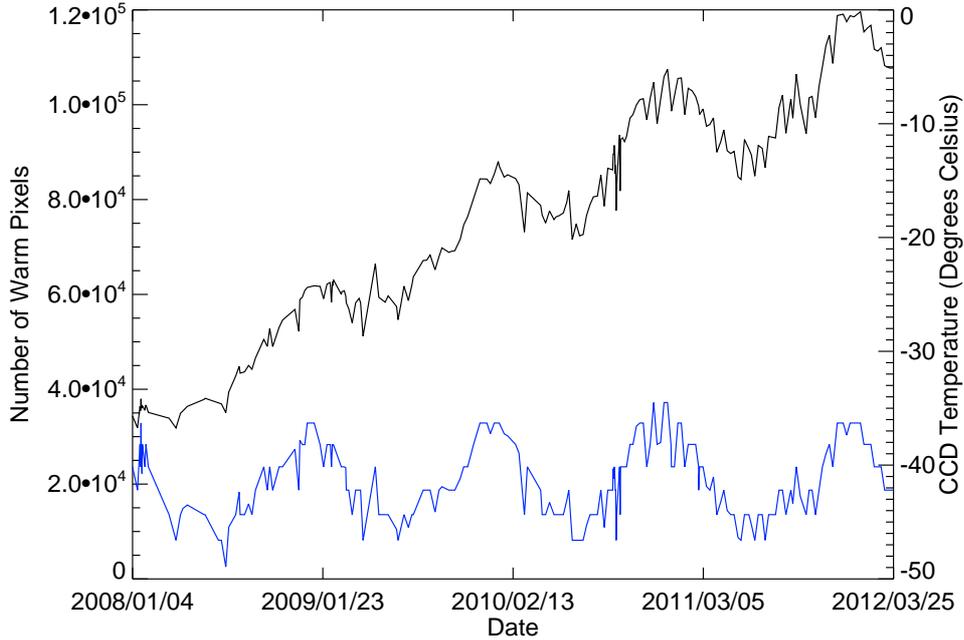}
   \caption{Number of warm pixels for the EIS CCDs. The (upper) black curve is the number of warm pixels and the (lower) blue curve is the corresponding CCD temperature.}
         \label{fig:walton1}
 \end{figure}

The hot and warm pixels are defects in the CCDs where rates of charge leakage are higher due to defects in the Si. 
A few exist in the as-manufactured device, and radiation damage accumulated over time adds to these. 
In the EIS orbit, the radiation effects are dominated by passages through the South Atlantic Anomaly (SAA). 
The hot and warm pixels are seen as spikes in the data which need to be calibrated out, \eg{} using $\sf eis\_prep$ (an IDL routine available in the EIS $\sf Solarsoft$ distribution). 
The positions of the warm and hot pixels are mapped and the information provided to $\sf eis\_prep$.
The distinction between hot and warm pixels is somewhat arbitrary. 
Pre-launch, hot pixels were defined as pixels where the room-temperature dark current rate was above $25\,000$ electron\,s$^{-1}$\,pixel$^{-1}$, 
for consistency with the criterion used by e2v in device screening and characterisation. 
In orbit, it has been found that radiation damage also causes pixels with lower charge-generation rates, but which are still above the CCD noise level, and so have to be taken account of.
 These are termed warm pixels, and the criterion used is that the dark signal is $>5 \sigma$ above the mean for a 100 second dark exposure. 
The hot and warm pixels generally follow the usual exponential temperature dependence of dark current in silicon, so cooling is effective in reducing the impact on the data. 
For EIS, the CCD temperature is higher than the pre-launch prediction of below  $-50\tc$, so the hot and warm pixels are greater in number. 
The XRT instrument on {\it Hinode} has the same type of CCDs as EIS but they operate at a much colder temperature ($\approx$--80$\tc$) and therefore do not have a problem with warm pixels.
The increase in hot pixels follows an approximately linear trend over time  and as of April 2012 the number of hot pixels per each CCD quadrant was $\approx$ 7800,
 which is equivalent to 1.5\,\% of the imaging area. The rate of increase of the warm pixels changes significantly and is temperature dependent, 
due to the dark-current variation with temperature (annual variations in the CCD temperature are due to aphelion and perihelion), as shown in  Figure \ref{fig:walton1}. 

As of March 2012, the average number of warm pixels for each CCD quadrant was $\approx$ 108 thousand (which is $\approx$ 21\,\% of the imaging area). 
The warm pixels are currently in a decreasing phase (CCD temperature is decreasing). 
The number of warm pixels in December of each year (hottest point) is increasing by $\approx$ 16 thousand per year. Based on this assumption the number of warm pixels will be:
\begin{itemize}
 \item $\approx$ 136 thousand (26\,\% of the imaging area) in December 2012,
 \item $\approx$ 152 thousand (29\,\% of the imaging area) in December 2013.
\end{itemize}

When a 30\,\% warm pixel level is reached the spectral-line-fit parameters will be affected (Private communication, P. Young 2012).
When the number of warm pixels (per CCD quadrant) reaches 157\,286 the 30\,\% level will have been reached. 
The highest number recorded so far was 120\,312 on 5 January 2012 (on the short wavelength CCD).

\subsection{Flat Field and EIS Sensitivity}
\begin{figure}
   \centering
   \includegraphics[width=1\textwidth, clip=false]{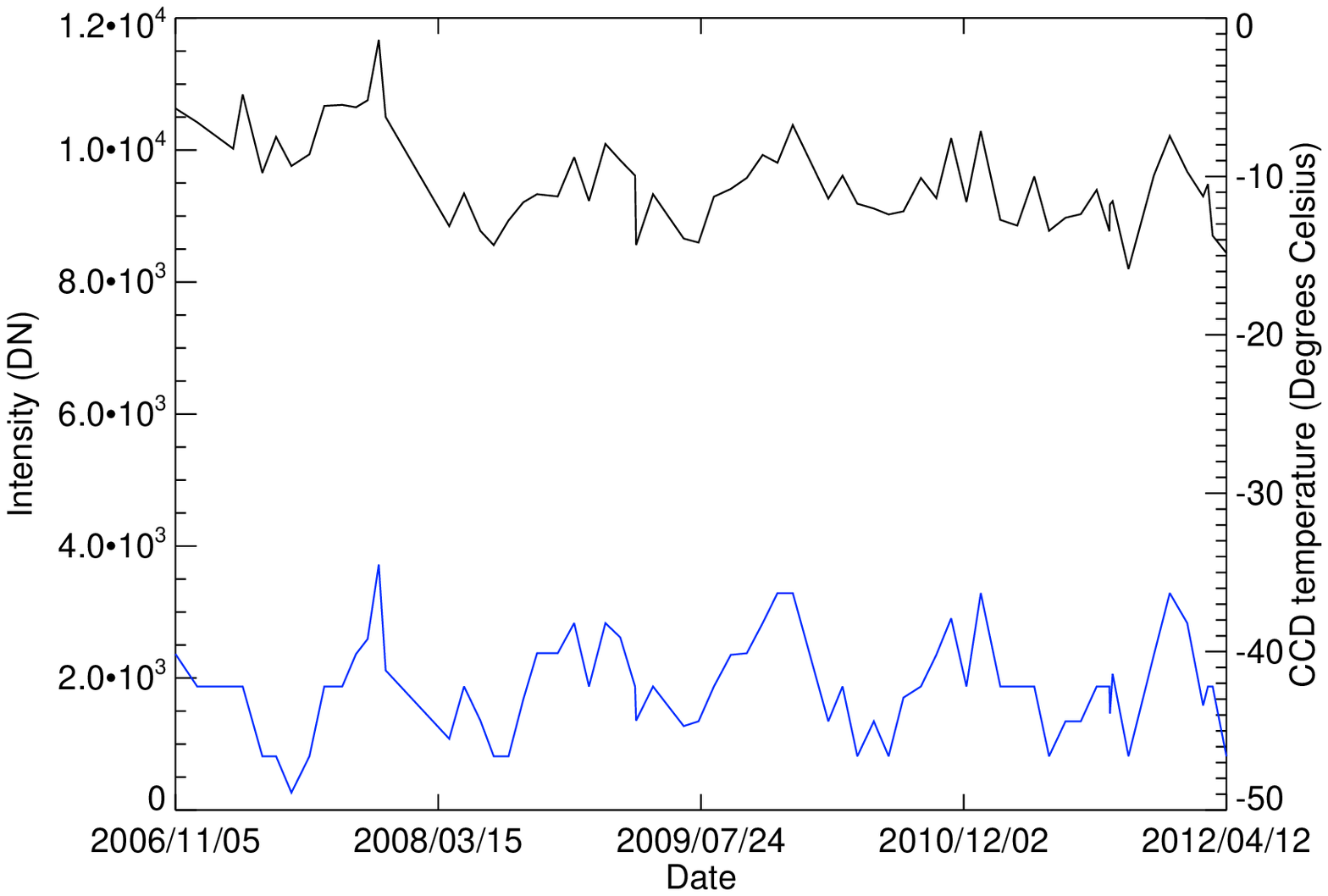}
   \caption{Flat-field intensity levels for the EIS CCDs using EIS LEDs (averaged over a 450$\times$450 pixel area). The (upper) black curve is the intensity level and the (lower) blue curve is the corresponding CCD temperature.}
         \label{fig:walton2}
 \end{figure}

The sensitivity levels for the EIS CCDs are monitored by using the data from LEDs flat-field images. 
The LEDs are blue, $\lambda_\textrm{peak}$ $\approx$ 430 nm, which approximately matches the absorption depth for the EUV photons.
The flat fields show that the intensity levels have not changed significantly since launch (Figure \ref{fig:walton2}), therefore indicating that any contamination is not on the CCDs. 

\begin{figure}
   \centering
   \includegraphics[width=1\textwidth]{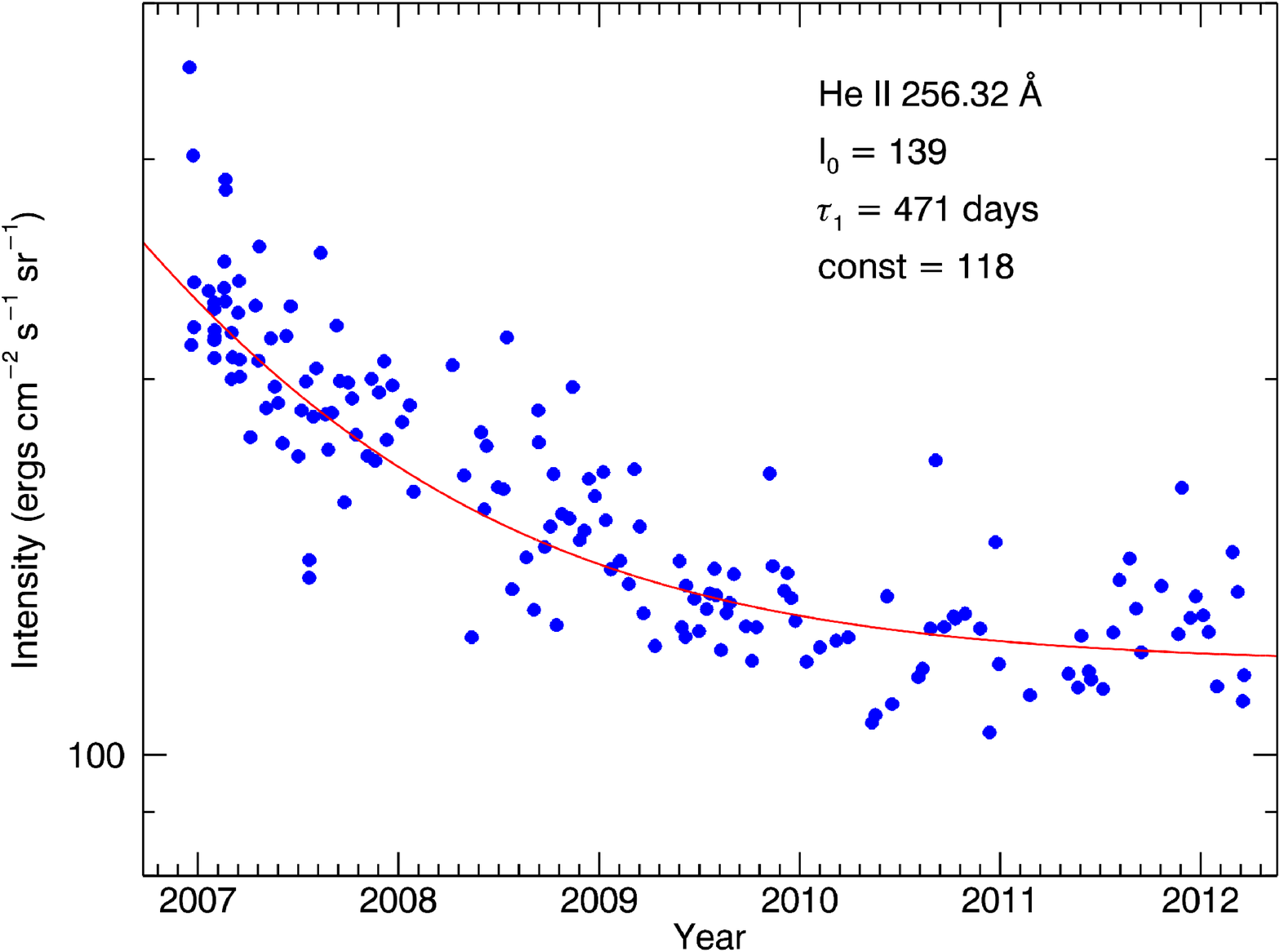}
   \caption{EIS sensitivity rate using \ion{He}{ii} observations (courtesy J. Mariska, NRL).}
         \label{fig:walton3}
 \end{figure}

A synoptic observation is run weekly, which observes a patch of quiet-Sun. 
The record of \ion{He}{ii} observations shows that the EIS sensitivity decay rate is slowing  (Figure \ref{fig:walton3}). 

The best-fit expression is now an exponential plus a constant. 
It is not yet clear whether some of the flattening is due to solar-cycle effects. 
Figures \ref{fig:walton2} and \ref{fig:walton3} indicate that the sensitivity decay is due to contamination/degradation of the optical elements rather than the CCDs.
The sensitivity changes are factored into the EIS analysis software ($\sf eis\_prep$).

The QCM (located at the entrance aperture) readings for EIS have been taken at weekly intervals since launch
 and provide an indication of the contamination levels of the critical instrument surfaces. 
The QCM data has shown a slight increase year to year which is to be expected. 
The QCM data from 2010 to 2011 saw a smaller increase compared to previous years. 
This agrees with the sensitivity measurements, which show that the decay rate is slowing ({\it i.e.} less contamination).

\subsection{CCD Annealing}
The plan so far has been to hold off from doing a CCD anneal or bake-out (heating the CCDs to around $+35 \tc$) for as long as possible, 
as it presents a small risk to the instrument (heaters require a high-power mode).
 Discussions at the recent EIS team meeting (April 2012) concluded with an annealing estimated for December 2013 which will be after seven years in orbit. 
Initially it was planned to perform a bake-out when the optical performance degraded but the flat-field intensity levels suggest that there is no contamination on the CCDs  (Figure \ref{fig:walton2}), and therefore a bake-out will not improve the optical performance. The current driver for an anneal is the number of warm pixels, and not contamination. 
The warm pixels will compromise the science when a 30\,\% level is reached.
The warm pixels are also impacting the EIS data compression; the effect being that the EIS telemetry allocation is reached as estimates are wrong (data compression degradation). 
The data-compression factors for EIS have been reduced by 10\,\% (five years into the mission). 
Until now, this reduction is only for the  {\it hot} season (October to April) -- when warm pixels are increasing. 
Data compression factors will be returned to nominal in the {\it cooler} season (April to October) when warm pixels are decreasing. 
It is hoped that, following bake-out, the data compression performance will revert to the post-launch level. 
The EIS operations team will have to experiment with anneal temperatures (+35 to $+45 \tc$) and durations to maximise recovery prospects of the hot and warm pixels.

\subsection{Conclusion}
The EIS instrument behaviour is nominal for a Low-Earth Orbit (LEO) space mission of its age. 
The CCDs are sustaining radiation damage as expected,
 although the amount of dark current per damaged pixel is higher than originally predicted due to the higher operating temperature (planned around $-60\tc$, achieved around $-45\tc$).
 The warm pixels will impact the science operations once their level reaches 30\,\% of the imaging area. 
At present they are mapped and removed via EIS processing software. It is hoped that a bake-out ($\approx +35\tc$) will recover most of the warm pixels. Bake-out is currently planned for the end of 2013.
The data compression performance was affected when the warm pixels reached a 25\,\% level (five years into the mission). 
EIS compression factors are now reduced by 10\,\% during the {\it hot} season. 
The optical performance degradation is better than expected (compared to similar missions and duration in orbit) -- it is expected that 1/e will be reached in seven years. 

\section{Long-term Stability of the Photometric Response of the STEREO/HI-1}
The twin {\it Solar TErrestrial RElations Observatory} (STEREO) spacecraft, which were launched in October 2006, are in heliocentric orbits at approximately 1 AU, with each spacecraft separating from the Earth by $22.5^\circ$ per year. 
STEREO-A is leading the Earth in its orbit, whilst STEREO-B is trailing the Earth.

The {\it Heliospheric Imager} (HI: \opencite{eyles2009}) instruments form part of the {\it Sun-Earth Connection Coronal and Heliospheric Investigation} (SECCHI: \opencite{howard2008}) suite of remote-sensing instruments aboard each of the STEREO spacecraft. 
They are primarily designed to observe coronal mass ejections (CMEs) as they propagate from the solar neighbourhood to Earth-like distances and beyond. 
Each HI instrument consists of two visible-light cameras, HI-1 and HI-2, having field-of-view (FOV) diameters of $20^\circ$ and $70^\circ$ respectively, 
and whose optical axes are aligned with the ecliptic plane in orientations providing an overall coverage of 4\,--\,$88.7^\circ$ solar elongation. 
The HI-1 and HI-2 telescopes consist of CCD cameras with fairly conventional transmission optics,
 and which are ``buried'' within complex baffle systems in order to provide the necessary high levels of solar stray-light rejection for imaging the faint emission from CMEs. 
The spectral band passes are 630\,--\,730 nm and 400\,--\,1000 nm for HI-1 and HI-2, respectively. The CCDs are passively cooled to below $-70\tc$.

\subsection{Initial Photometric Calibration} 

The initial photometric calibration of the HI-1 telescopes \cite{bewsher2010} was based on data from the start of STEREO mission operations up to December 2008.
 The intensities of stars with R magnitudes $\leq 12$ and within 100 pixels radius from the center of the FOV were measured using aperture photometry. 
The stars and their spectral types were identified from the NOMAD catalogue \cite{Zacharias2004}. 
Predicted intensities were calculated by folding standard stellar spectra {\it S($\lambda$)} \cite{pickles1998} through an optimised model of the instrument response function.
The model of the instrument response was optimised using all available pre-flight calibration data, CCD and optics manufacturers response specifications, etc.

\begin{figure}
               \includegraphics[width=1\textwidth,clip=]{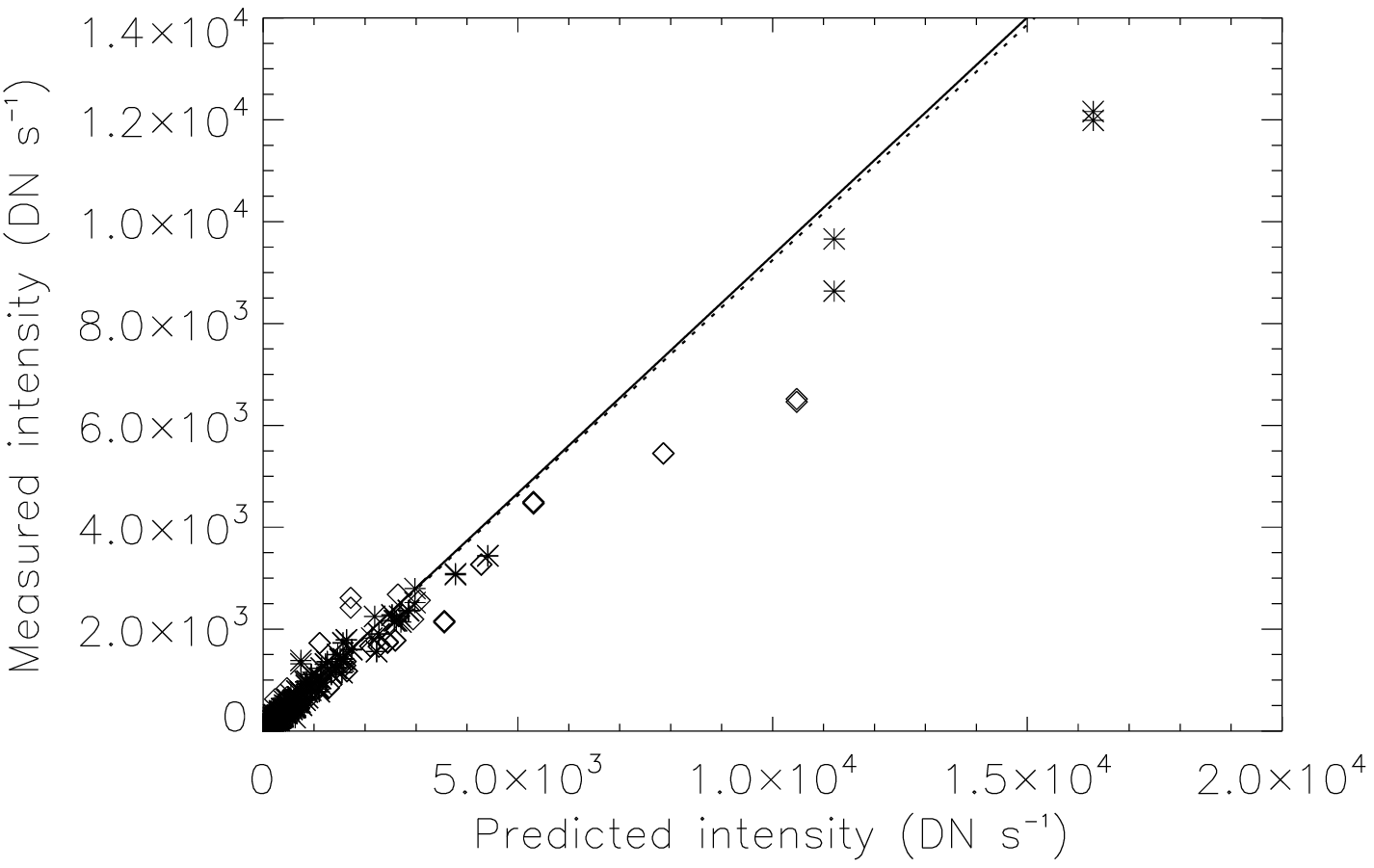}\\
               \includegraphics[width=1\textwidth,clip=]{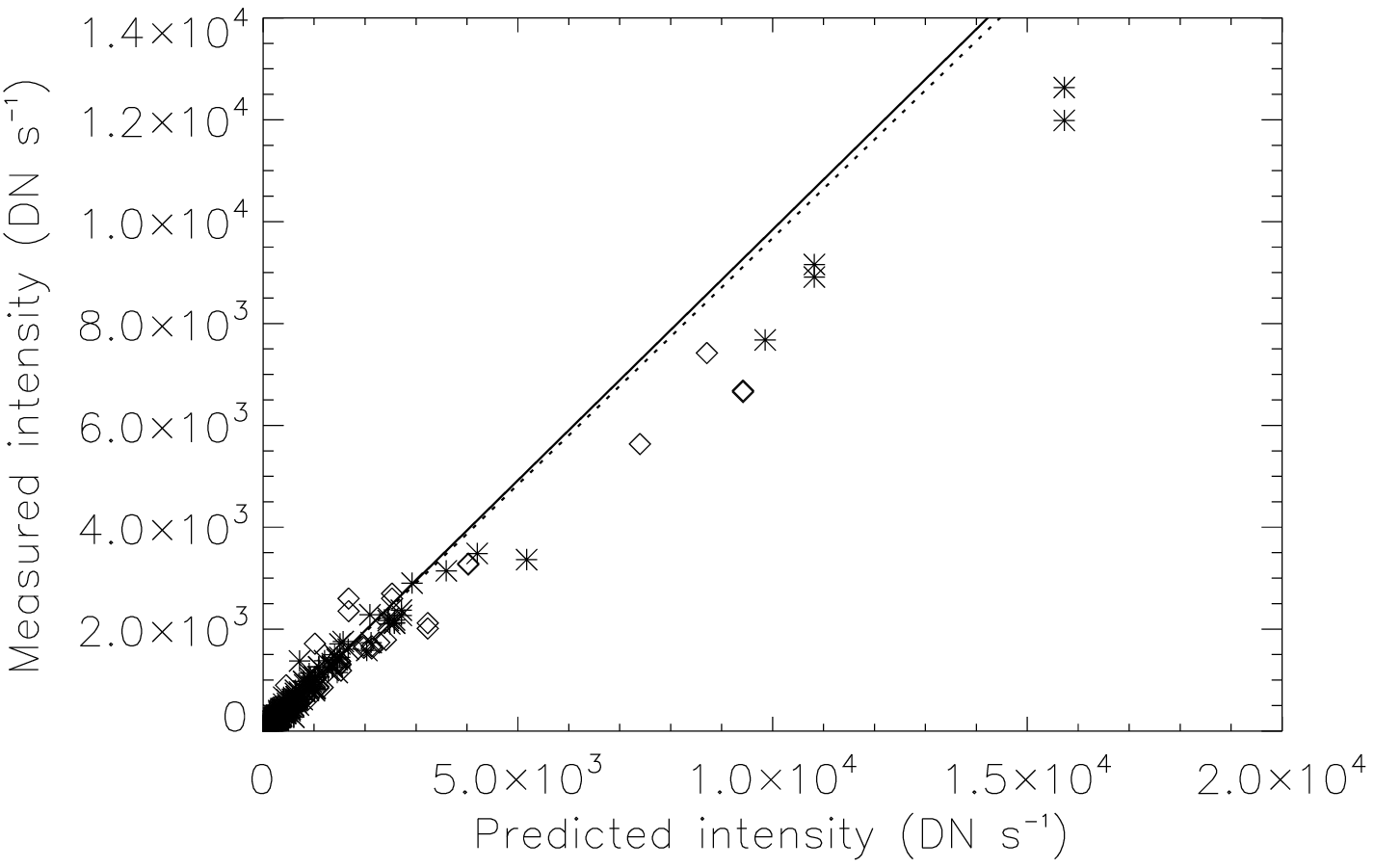}     
\caption{Measured {\it versus} predicted intensities (in DN\,s$^{-1}$) for a large number of stars in the HI-1A (top) and HI-1B (bottom) initial photometric calibrations. 
Updated from \protect\inlinecite{bewsher2010}.}
   \label{fig:eyles1}
   \end{figure}

Figure \ref{fig:eyles1} shows the measured {\it versus} predicted intensities for large populations of stars. 
Apart from a few outliers at high intensities (due to detector saturation effects), the stars lie close to a fitted straight line of slope $\mu$, i.e. $C_\mathrm{measured}$ = $\mu C_\mathrm{predicted}$.\\
The photometric calibration factor [$\mu$] represents an overall normalisation error in the instrument response, the value $\mu=1$ representing a perfect calibration.
The values obtained for $\mu$ were 0.93 and 0.99 for HI-1A and B respectively, with the total number of stars fitted being 903 and 541. 
No significant differences in $\mu$ were found according to spectral type, confirming the validity of the model of the instrument spectral response.

\subsection{Stability of the Photometric Response} 
In order to evaluate the stability of the photometric response, the above analysis was performed separately for each of the complete orbits of the STEREO spacecraft.
The derived values for $\mu$ are shown in Table \ref{T-HI1}. The values for orbits 1\,--\,4 relative to the background star field are taken from \inlinecite{bewsher2012},
and the values for orbit 5 and the overall value of $\mu$ are new to this article.
As previously, $\mu$ was determined using stars within 100 pixels radius from the center of the FOV \cite{bewsher2012}. 
Complete orbits were used in order to avoid any possible systematic effects due to different star populations being used in each case.

\begin{table}
\caption{The photometric calibration factor [$\mu$], and the number of stars used to determine $\mu$.}
\label{T-HI1}
\begin{tabular} {lcccccc}                    
  \hline     
  &Orbit 1 &Orbit 2&Orbit 3&Orbit 4& Orbit 5& Overall\\
  \hline
HI-1A& 0.926& 0.933 & 0.927 & 0.930 & 0.935& 0.930 \\
No. stars & 430 & 418 & 433 & 424 & 428& 2141\\
HI-1B& 0.998& 0.987 & 0.989 & 0.993 & 0.981 & 0.989 \\
No. stars & 387 & 384 & 455 & 417 & 368 & 2029\\
\hline
\end{tabular}
\end{table}

It is clear from Table \ref{T-HI1} that there are no systematic changes in the photometric response of either instrument at a level of 1\,\% or better. 
The analysis was also repeated for various selected regions of the FOV and again the response was found to be stable to 1\,\% or better, 
although some variations at the level of 2 \,--\, 3\,\% in the value of  $\mu$  for different regions of the FOV were found, 
indicating some small systematic errors in the HI-1 flat-field corrections \cite{bewsher2012}.

\subsection{Conclusions} 
We have shown that from the start of mission science operations until the end of the fifth orbit of the two spacecraft relative to the background star field
(5 Dec 2011 and 18 July 2012 for STEREO A and B, respectively), the photometric response of the HI-1 cameras has remained stable to 1\,\% or better. 
This is significantly better than the long-term stability of the white-light coronagraphs onboard the SOHO mission,
 where \inlinecite{thernisien2006} found a decrease of sensitivity of the {\it Large Angle and Spectrometric Coronagraph} (LASCO) C3 instrument by 3.5\,\% over eight years of operation, 
whilst \inlinecite{llebaria2006} reported a degradation of the LASCO-C2 instrument by 0.7\,\% per year. 
Whilst white-light instruments are not as sensitive to degradation as UV instruments, 
the excellent stability of the HI instruments vindicates the extensive precautions taken during their design and development. 

\section{SOLSPEC: A Spectrometer With Onboard Control of Aging}

The {\it SOLar SPECtrum} (SOLSPEC) is a spectrometer, that flew several times on the Space Shuttle, and its twin instrument was placed on the {\it European Retrievable Carrier} (EURECA) platform for ten months.
The Shuttle flight has gathered data to build the {\it Atmospheric Laboratory for Applications and Science} (ATLAS) 1 and 3 spectra \cite{thuillier2009}, 
which comprised the Upper {\it Atmospheric Research Satellite} (UARS) / {\it Solar Ultraviolet Spectral Irradiance Monitor} (SUSIM) and {\it Solar Stellar Irradiance Comparison Experiment} (SOLSTICE) data from lyman-$\alpha$ (121.6 nm) to 200 nm, and ATLAS-{\it Shuttle Solar Backscatter Ultraviolet} (SSBUV), SUSIM, and SOLSPEC from 200 to 400 nm,
 ATLAS-SOLSPEC from 400 to 850 nm, and EURECA- {\it SOlar SPectrum} (SOSP) from 800 to 2400 nm. 
The ATLAS spectra are calibrated into the absolute radiometric scale of the black-body radiator of the Observatory of Heidelberg, NIST standards spectral irradiance (tungsten and deuterium lamps). 
The SOLSPEC instrument has been upgraded for operations onboard the {\it International Space Station} (ISS) by implementing several changes given the lessons learned from the previous missions and by adding several new components in order to provide an instrument able to operate for several years in the space environment.
SOLSPEC has now been in operation on board the ISS since February 2008.
To cover the 165\,--\,3080 nm range, three double spectrometers are used which are equipped with concave holographic gratings made by Jobin--Yvon. 
By rotating the six gratings mounted on the same mechanical axis, the range 165\,--\,3080 nm is scanned in ten minutes with a mechanical precision corresponding to 0.01 to 0.1 nm, from the UV to the IR channels.
 To reduce the flat-field effect, diffusers are placed between the entrance pre-slit and spectrometer first slit.
As the ISS environment could not always be clean in terms of contamination, and as the diffusers could degrade by EUV solar radiation,
 two wheels each carrying a hole and two quartz plates can be placed alternatively in front of the entrance pre-slit.
 These plates allow protection of the entrance slits from contaminants deposition, which can be ultimately polymerized by the solar EUV. 
In that case, the quartz-plate transmission decreases; however, it can be measured in-flight by using the ratio of solar observations with and without the quartz plate in front of the entrance
 slit. One plate is mainly used for each observation. Using the Sun, its transmission is compared to the not frequently used quartz-plate transmission. 
For each spectrometer, a wheel is equipped with second-order and/or neutral filters. 
The latter are used to reduce the signal given by the instrument responsivity and the solar irradiance variation with wavelength.

\subsection{Pre-Flight Absolute Calibration}

SOLSPEC has been calibrated at the PTB using a black-body radiator. 
One of these black-body sources (BB3200pg) represents the primary standard for the realization of the spectral irradiance scale \cite{sperfeld1998}.
 Taking into account the distance between the black-body source and the SOLSPEC entrance slit, the size of the entrance slit, the black-body aperture, and the black-body temperature, 
the spectral irradiance is calculated for any given wavelength using the Planck law. The black-body cavity temperature is around 3000\,K and it is known to within 0.44\,K. 
The black-body emission being calculated, and its ratio to the count number recorded by SOLSPEC, allows the solar signal to be converted to absolute irradiance. 
Below 200 nm, the black-body source does not generate enough signal, and hence deuterium ($\textrm{D}_\textrm{2}$) lamps provided by the PTB were used. During the calibration campaign, several spectra using the internal lamps were recorded as a reference in the relative scale.
For the whole spectral range, the accuracy of SOLSPEC stays within 3\,\%.
The SOLSPEC instrument is described in detail by \inlinecite{thuillier2009}.

\subsection{Onboard Calibration Means and Instrument Degradation in Space}
$\textrm{D}_\textrm{2}$ and tungsten ribbon lamps are used for checking the instrument stability with time. 
The light from these sources is carried by using optical fibres, mirrors, and lenses.
 The relationship between the gratings mechanical position and wavelength is measured by using a hollow cathode (HC) lamp filled with argon (Ar) delivering lines in the UV,
 visible, and near IR. These lines also allow the measurement of the instrument slit function and the dispersion law (relationship between the grating position and wavelength). 

Degradation in visible and IR domains is of about a few percent, and can be corrected by measuring the transmission of the quartz plates and by using the internal ribbon tungsten lamps. As expected, the degradation in the UV is significant.

\begin{figure}
   \centering
   \includegraphics[width=1\textwidth]{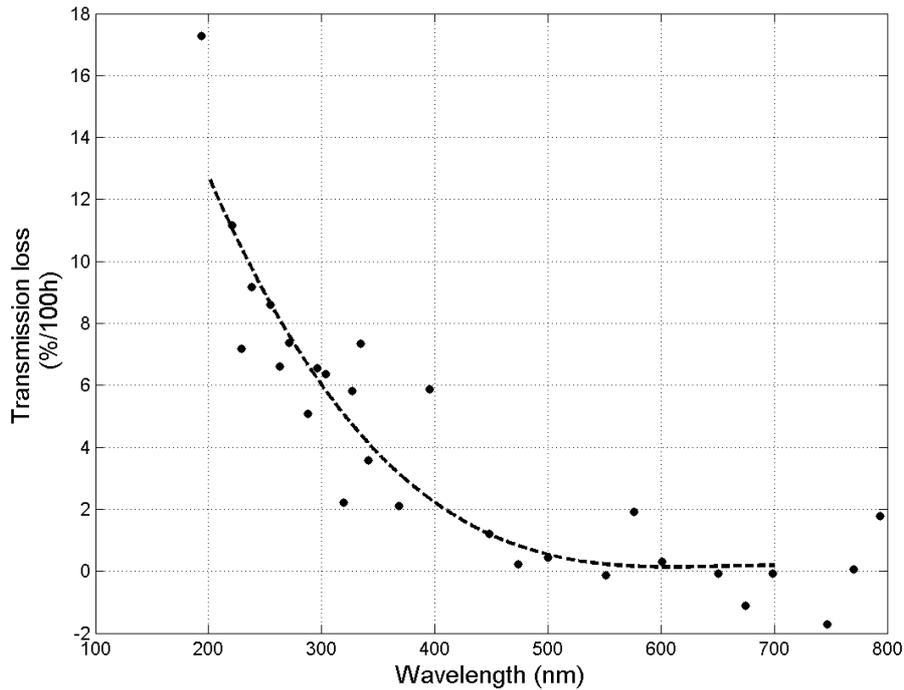}
   \caption{Transmission degradation of the quartz plate used for the UV--VIS solar spectrum measurements. The transmission is measured by comparing the signal with and without the quartz plate.}
         \label{fig:solspec1}
 \end{figure}

Figure \ref{fig:solspec1} shows the transmission loss of the most frequently used quartz plate.  
The instrument responsivity change is derived from comparison of the transmission of the second quartz plate (infrequently used), direct quartz-plate transmission measurements, and $\textrm{D}_\textrm{2}$ lamp data. 
After the $\textrm{D}_\textrm{2}$ lamp power-supply failure, we used the HC lamp lines intensity based on the following principle:

\begin{itemize}
\item In the laboratory, as well as in orbit, we have verified that the lines emitted by the HC lamp have a covariance intensity change in time. 
In other words, the ratio of line intensity at two different wavelengths is constant, {\it i.e.} independent of the line intensity, which slowly decreases likely due to a gas leakage. 

\item As there is a corrected aging (as explained above), in the visible (or IR) spectrometer, 
the HC-lamp lines intensity can be corrected and the percentage of correction is in particular applied to the UV lines providing data to correct the UV spectrometer responsivity. 
\end{itemize}

\subsection{Conclusion}google
The SOLSPEC design has been validated by the {\it SpaceLab I}, ATLAS 1, 2, 3 and EURECA missions. However, the duration of these missions were about one week except for EURECA, which lasted ten months.  
For a mission aiming to operate for several years, it was necessary to design an instrument having its own capability to provide data for aging corrections given its location in space. 

\section{Solar Instruments Onboard PROBA2}

The {\it Project for On Board Autonomy} (PROBA) satellites are part of ESA's In orbit Technology Demonstration Programme, {\it i.e.} missions dedicated to the
demonstration of innovative technologies through small satellites.
On 2 November 2009, PROBA2 was launched into a Sun-synchronous polar orbit allowing quasi-permanent solar observation. 
Two solar observation experiments, the {\it Sun Watcher with Active Pixels and Image Processing} (SWAP:  \opencite{Seaton2012};
\opencite{Halain2012}) and the {\it Large-Yield RAdiometer} (LYRA: \opencite{Hochedez2006}) on PROBA2 are test platforms for new technologies. 

The absolute radiometric response of the two instruments, through collaboration between the Max-Planck-Institut f\"ur Sonnensystemforschung (MPS) and the PTB,
has been established before flight at the primary radiometric source standard,
the synchrotron radiation beamline of PTB at the {\it Berlin Storage Ring for Synchrotron Radiation II} (BESSY II).

\subsection{A Characterization of SWAP Degradation}

SWAP is a single-band EUV telescope which observes the solar corona in a passband centered on 17.4 nm and with a 54$\times$54 arcmin FOV, 
has a novel off-axis Ritchey--Chretien design with two mirrors with multilayer coatings for EUV reflectivity and a complementary metal-oxide-semiconductor (CMOS) active pixel sensor (APS). 
A scintillator coating (P43) converts EUV photons into visible photons to which the detector is sensitive. 
Spectral selection is achieved by the combination of multilayer coatings and two Al-foil filters, one of which is located at the entrance aperture and the other in front of the focal-plane assembly. 
SWAP has operated essentially continuously since shortly after PROBA2's injection into its polar Sun-synchronous orbit in November 2009 at an approximate altitude of 725 km. 
Additionally, since SWAP does not have a door or shutter, its optical and electronic systems are continuously exposed to EUV input from the corona.
We characterize SWAP's degradation in three ways: first, we compare the total response of the instrument to well-calibrated spectral measurements from the {\it Extreme Ultraviolet Variability Experiment} (EVE: \opencite{eve}; \opencite{esp}) onboard the{\it Solar Dynamics Observatory} (SDO) spacecraft; 
second, we measure the number of improperly performing pixels in SWAP's 1024$\times$1024 pixel CMOS detector; and finally, we roughly measure the evolution of SWAP's flat-field using a set of onboard LEDs.
Further information on the ground-based calibration from PTB/BESSY II is discussed by \inlinecite{Seaton2012}, and the in-flight calibration by \inlinecite{Halain2012}.

\subsubsection{Spectral Response}

Since it is not possible to obtain in-orbit EUV images of a standard and well-calibrated source, degradation in the SWAP response function must be measured indirectly. 
In order to do this, we compared the mean SWAP response per pixel for solar images obtained regularly throughout the mission to corresponding spectra from EVE. 
To achieve this comparison, EVE spectra are first converted from units of total flux per wavelength to photon flux, then modulated by the laboratory-measured SWAP response function,
 and integrated across SWAP's entire bandpass. This procedure yields a single value with units of DN\,s$^{-1}$\,pixel$^{-1}$ that we can compare to the mean instrumental response
 in images obtained at the same time as the corresponding spectrum.  

 \begin{figure}
  \centering
  \includegraphics[width=1\textwidth]{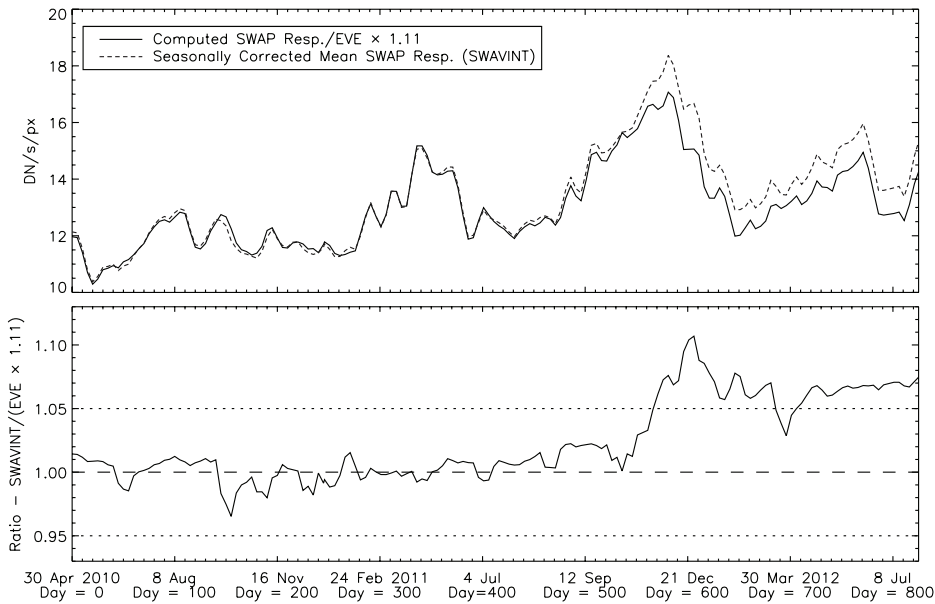}
  \caption{Comparison of SWAP measured intensity (dashed) and SDO/EVE computed intensity (solid). 
The top panel shows the variation of intensity with time due to changes in solar irradiance near 17.4 nm, the lower panel shows the ratio of the two values over time. 
Adapted from \protect\inlinecite{Halain2012}.}
         \label{fig:swap1}
  \end{figure}

We then compare the evolution of these two values over time to assess the rate of degradation of SWAP's overall response over the course of the mission. 
Note that a multiplicative factor of 1.11 is applied to the EVE-based curve before our analysis. 
 The need for such a factor is probably the result of a combination of uncorrected degradation of EVE in orbit and error in measuring the SWAP response in the laboratory; 
the factor to be used was measured empirically, and could be the result of many different contributions.
 Figure~\ref{fig:swap1} shows this comparison; although the two curves are closely correlated,
 driven by the variation in coronal irradiance near 17.4 nm, they diverge at the end of 2011, the reason for which is still unclear.

\subsubsection{Detector Degradation}

\begin{figure}
  \centering
 \includegraphics[width=1\textwidth]{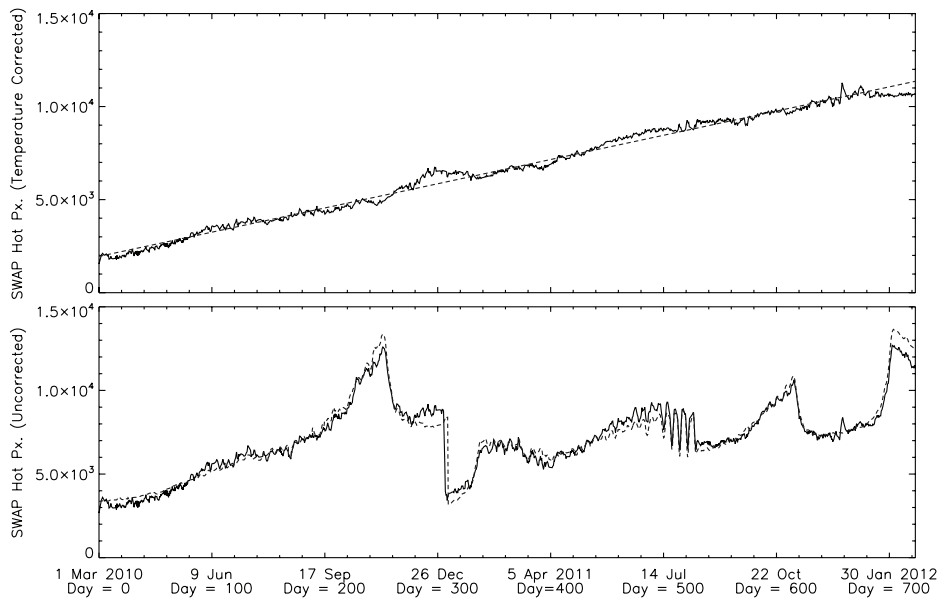}
  \caption{Spikes detected in SWAP images {\it versus} time. The top panel shows the rate of increase in detections with the effects of temperature variation removed. 
The lower panel shows the variation including temperature effects. The dotted curve in both panels shows the effect of a linear increase in spikes with a rate of 13 spikes per day. 
In the lower panel this has been adjusted to reflect temperature variation as well, showing that the linear increase is indeed a good match for actual detector behavior.
Adapted from \protect\inlinecite{Halain2012}.}
         \label{fig:swap2}
  \end{figure}

A potentially more significant problem for SWAP is the breakdown of electronic components, especially those associated with its CMOS--APS detector,
 which is the first of its kind used for an EUV solar telescope. A complete discussion of SWAP's detector and its on-ground performance testing was given by \inlinecite{DeGroof2008}. 
Unlike CCD detectors, which have been used in nearly all solar imaging missions for decades, each pixel in a CMOS--APS detector has its own analog (amplifier) readout electronics, 
so a failure of these electronics can render only individual pixels inoperable.
We monitor detector performance primarily by tracking the number of hot pixels removed by the de-spiking routine in SWAP's image-calibration software. 
This value is strongly influenced by the evolution of detector temperature, so to separate changes in detector performance from the thermal evolution of SWAP, 
we model the temperature dependence of hot pixels using an empirically determined polynomial model and normalize the evolution of spikes in time using this model. 
Figure \ref{fig:swap2} shows both the normalized and unnormalized curves for a period of about 700 days after the end of PROBA2's commissioning period.
In the normalized plot, the number of spikes clearly increases linearly in time at a rate of about 13 spikes day$^{-1}$ or about 4800 spikes year$^{-1}$. 
This corresponds to a loss of only about 0.45\,\% of all detector pixels per year. 
We note that this linear increase in pixel damage is consistent with the results ground-based radiation exposure tests described in the detector datasheet produced by Cypress (formerly FillFactory), the detector producer.
The large jump around day 300 (Figure \ref{fig:swap2} lower panel) is due to a refinement in SWAP calibration procedures. 
The two dips in the number of spikes near December and January of 2010 and December 2011 are the result of PROBA2's reduced operating temperature during the spacecraft's eclipse season.

\subsubsection{Image Quality Degradation}
The final type of degradation that can affect SWAP is the degradation of intrinsic image quality due to losses of efficiency in the optical components as a result of EUV ``burn-in'' or deposition of contaminants on optical surfaces. 
This type of degradation is the most difficult to measure, since, as is the case for all space-based EUV telescopes, there is no standard EUV source available with which to measure SWAP's flat-field while in flight. 
However, this type of degradation was the principal cause of loss of quality in images from EIT on SOHO \cite{Clette2002} so it is worthwhile to study its role in SWAP degradation to the extent that we can do so.
Although the production of a true gain calibration for SWAP is very difficult, SWAP carries two visible LEDs that can help to reveal strong variations in image quality. 
For this analysis we compare the ratios of LED brightness at the beginning of PROBA2's mission with more recent observations of the LEDs. 
Since the LEDs are located close to SWAP's focal plane assembly (FPA), we cannot characterize any changes in filter or mirror performance with this measurement. 
However, any significant changes in the performance of the optical path would likely show up in our analysis of SWAP's spectral response. 
Nonetheless, it is worth pointing out that this analysis applies only to the FPA components and LEDs themselves.
We compared LED images obtained as part of a bi-weekly calibration campaign performed throughout the mission to study changes in image quality over time. 
By computing the pixel-by-pixel ratio of LED images from the beginning of the mission to LED images from the end of the mission, we can determine whether any spatially coherent degradation has occurred. 
Comparing images from April 2012 to LED images from the early-mission commissioning phase revealed a small,
ring-shaped decrease in detector response, roughly coincident with the location of the solar limb, where the brightest coronal emission occurs, in nominal SWAP images. 
This suggests that there has been some degree of burn-in over the course of the mission. However, the decrease was only a few DN per pixel,
which corresponds to a net decrease in instrumental response of less than 0.1\,\% of total signal in well-exposed images.
Since this level is far below instrumental noise levels and, as a result of image compression, is in fact undetectable in 
nominal science images, we conclude that this type of degradation is not a significant concern for SWAP.
Additional discussion of the use of LED images to diagnose SWAP degradation, including a figure that shows this effect, is given by \inlinecite{Halain2012}.

\subsubsection{Conclusion}
SWAP has a dual role; it is both a scientific instrument and a test platform for new technology. 
While many such space-based EUV instruments have experienced significant degradation during the initial years in orbit, SWAP has shown itself to be remarkably robust against degradation of any kind. 
This analysis suggests SWAP has only experienced one significant type of degradation: failures in the detector electronics, which have occurred at a rate of less than 5000 out of approximately 10$^6$ pixels (less than 0.5\,\%) every year. 
The lessons learned from SWAP's simple, efficient, and robust design are especially applicable to instruments intended primarily for space weather monitoring such as the proposed {\it EUV Solar Imager for Operations} (ESIO) instrument. 
Such instruments, which often are expected to operate with limited resources, must be long-lived and dependable and thus must be highly robust against adverse conditions in the space environment to which they will be exposed.

\subsection{LYRA Degradation after Two Years in Orbit}
The {\it Large-Yield RAdiometer} (LYRA), observes the Sun in four spectral bands that range from UV to soft X-ray.
It consists of three units that are redundant but not technically identical.
 While each unit consists of the same four spectral channels covering a wide emission temperature range, these channels are realized by different filter--detector combinations. 
Three types of detectors were used, conventional Si photodiode detectors (AXUV type from IRD), as well as two types of diamond detectors,
 which have the advantage of being radiation resistant and insensitive to visible light \cite{BenMoussa2006}. 
Another advantage of LYRA is its high observation cadence, up to 100 Hz.
 LYRA uses two calibration LEDs per detector to individually monitor the possible detector degradation over the mission lifetime. 

LYRA channel 1  (lyman-$\alpha$) covers a narrow band around 120\,--\,123 nm, plus, unfortunately, 
a major contamination caused by longer wavelengths. LYRA channel 2 (herzberg) covers the interval 190\,--\,222 nm in the Herzberg continuum.
 LYRA channel 3 (aluminum) covers the 17\,--\,70 nm Al filter range including the strong \ion{He}{ii} 30.4 nm line, as well as soft X-ray contribution below 5 nm. 
LYRA channel 4 (zirconium) covers the 6\,--\,20 nm Zr filter range with the highest solar variability, as well as soft X-ray contribution below 2 nm. 
For a detailed description of the mission, see \inlinecite{Hochedez2006}.

Since PROBA2 is a combined science and technology mission, the goal was not only to provide scientific data for the study of solar flares, 
space weather, and aeronomy, but also to observe the performance of new technologies in space. 
Thus, the observation of the instrument's degradation and its causes is an important part of the mission goal.

\subsubsection{Calibration}

The spectral response of the twelve possible filter--detector combinations was tested before launch with a standard source (PTB/BESSY II); for details see \inlinecite{BenMoussa2009a}. 
The various nominal spectral intervals were defined accordingly, such that they cover at least 95\,\% of the response. 
This does not apply for the lyman-$\alpha$ channel, since the unwanted longer-wavelength contributions to this channel depend on the detector technology.

LYRA is continuously observing the Sun, basically with its unit 2, for more than two years. 
The data presented on the PROBA2 website (\url{proba2.sidc.be}) include daily plots, three-day plots, monthly overviews, flare lists, 
and a comparison with the soft X-ray of NASA's Geostationary Operational Environmental Satellites (GOES) satellites.
 LYRA data are available in daily FITS files; users can chose between uncalibrated and calibrated time series in full temporal resolution, and calibrated data averaged in one-minute intervals.

\subsubsection{Degradation}

\begin{figure}
   \centering
   \includegraphics[width=1\textwidth]{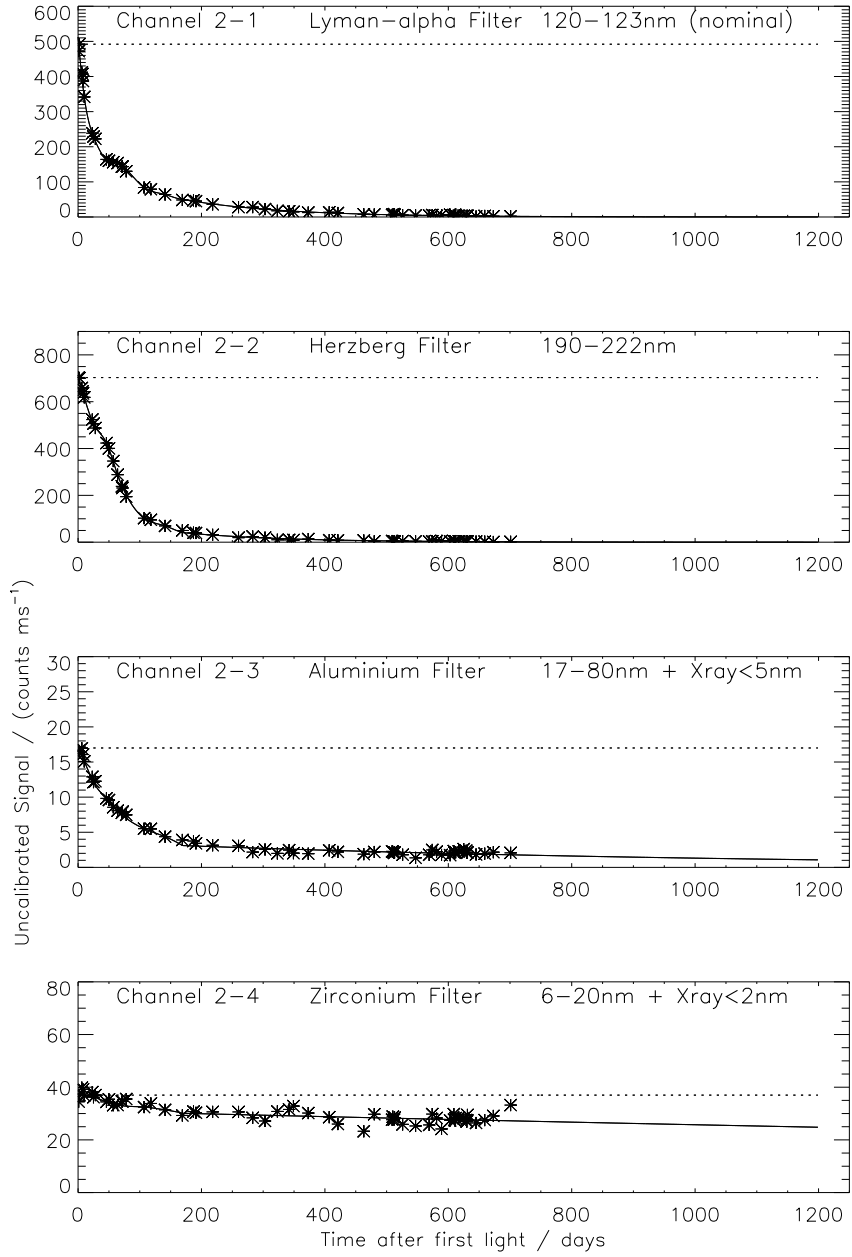}
   \caption{Temporal degradation and estimation of future trends for LYRA unit 2.}
         \label{fig:lyra1}
 \end{figure}

LYRA unit 2 consists entirely of new diamond detectors to be tested in space.
It is therefore selected to be the ``nominal" unit and ihas been used, almost without interruption, since 6 January 2010, the day that LYRA first opened its covers. 
Unit 1 and unit 3 are only opened and used sporadically, for limited campaigns and for calibration purposes. 
Consequently, unit 2 degraded quite fast, especially in its longer-wavelength channels; see Figure \ref{fig:lyra1}. 

LYRA's original spectral response, as measured in the laboratory, can only be compared in a reasonable way to other space instruments when data of the first-light day are used, {\it i.e.}, 
before heavy degradation set in.
It was thus decided to calibrate LYRA with the help of a combined solar spectrum observed on 6 January 2010 by SOLSTICE onboard the {\it Solar Radiation and Climate } (SORCE) 
and the SEE (Solar EUV Experiment) on the TIMED mission.
 The calibration is then extended by adding the estimated loss by temporal degradation. 
This method has the advantage of leaving the flare components within the shorter-wavelength channels untouched, since it is observed that these signals do not degrade.
 In addition, the shorter-wavelength channels react more to long-term solar variability;
 therefore this variability has been adjusted with the help of ch3-4 ({\it i.e.} unit 3, channel 4), the zirconium channel of unit 3, which is assumed to be non-degrading.
 
\begin{figure}    
   \centerline{
               \includegraphics[width=0.5\textwidth,clip=]{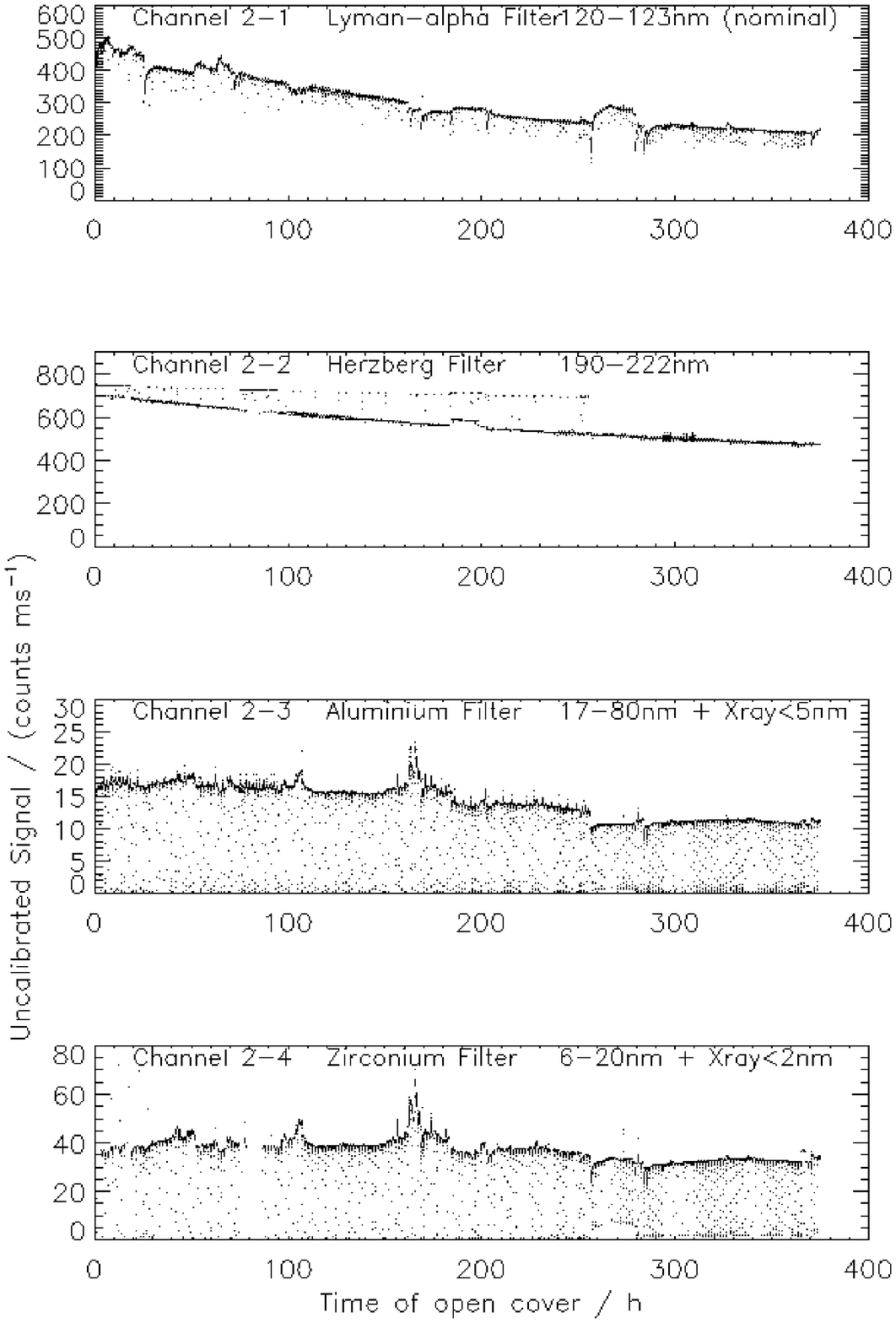}
               \includegraphics[width=0.5\textwidth,clip=]{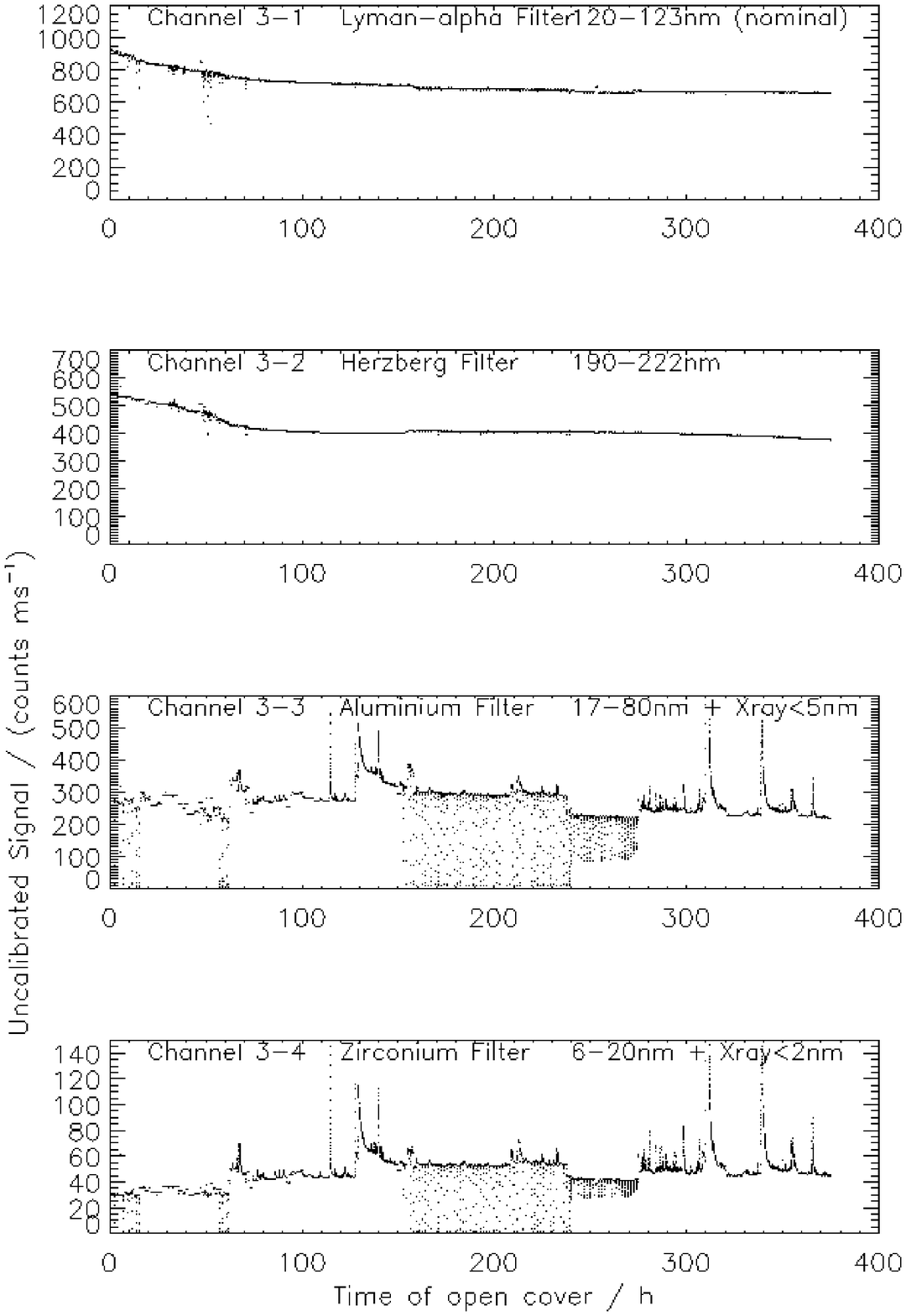}}
\caption{ Comparison of the first 375 hours of unit 2 (quasi-continuous 6 January 2010\,--\, 5 February 2010) and unit 3 (campaigns 6 January 2010 \,--\, 15 March 2012).}
   \label{fig:lyra2}
   \end{figure}

By mid-March 2012, unit 3 had been open to the Sun for approximately 375 hours. 
It could thus be compared with the first 375 hours of open unit 2, which was reached around 05 February 2010, within the commissioning phase.
 The result is shown in Figure \ref{fig:lyra2} and Table \ref{T-lyra}.

\begin{table}
\caption{Relative signal losses (to the output signal on 6 Jan. 2010: first light) in the four LYRA channels of unit 2 and unit 3, each after 375 hours of open covers (dark currents subtracted).
 To remove the solar variation contribution in the shorter wavelength(*), ch2-3, ch2-4 and ch3-3 were adjusted relatively to ch3-4, which is assumed to be non-degrading.}
\label{T-lyra} 
\begin{tabular} {cccc}                    
  \hline     
LYRA-unit2 &  degradation [\%]& LYRA-unit3 &  degradation [\%]\\
  \hline
ch2-1 & 58.3& ch3-1 & 28. 3\\
ch2-2 & 32.5&ch3-2 & 30.9 \\
ch2-3 & 28.7* & ch3-3 & 45.2* \\
ch2-4 & 10* & ch3-4 & 0*\\
\hline
\end{tabular}
\end{table}

The loss percentage for the short-wavelength channels is calculated under the assumption that the solar irradiance had dropped by 2\,\% within the period covered by unit 2,
 and that it had increased by 55\,\% in the period covered by unit 3; this is observed by LYRA ch3-4 (unit 3, channel 4). 
It is furthermore assumed that the solar variation as reflected in ch2-3, ch2-4, ch3-3, and ch3-4 is linearly dependent, and that ch3-4 is not degraded.

The loss in ch3-1, a channel which has significant contributions from visible and IR radiations (Si detector), appears to be smaller than in ch2-1, 
which only has a significant contribution from UV, apart from the Lyman-$\alpha$ line. Ch3-3, after removal of the solar variation,
 appears to degrade initially as fast as the longer-wavelength channels, while channel 4 appears to degrade slower than the others. 
Meanwhile, it is observed by regular calibration campaigns -- using the LEDs with covers closed --  that the photodetectors made on diamond do not show any degradation 
while the Si AXUV detectors a slight increase of its dark current.

As can be seen in the unit-3 part of Figure \ref{fig:lyra2}, the degradation slows down after sixty hours,
which corresponds to campaigns around the end of 2010, {\it i.e.} after approximately one year of operations. 

\subsubsection{Comparison with EURECA/SOVA}

Taking these results into account, it appears that the degradation of LYRA is basically not due to detector loss, but it is due to molecular contamination on the front optical-filter surface.
It is interesting to compare this to another space instrument, {\it Solar Oscillation and Variability} (SOVA) onboard EURECA, which experienced both kinds of losses.

The SOVA experiment has three channels at around 335 nm (UV), 500 nm (visible), and 862 nm (NIR).
 Its {\it sunphotometers} (SPM) were operated in space for eleven months onboard EURECA. 
SOVA was launched and retrieved with shuttles in 1992/93, and inspected at the Physikalisch-Meteorologisches Observatorium Davos/World Radiation Center after retrieval in 1994.
They found a yellow--brownish stain of unknown composition on the quartz windows and the apertures. For a more detailed description, see \inlinecite{Wehrli1996}. 

\begin{figure}
   \centering
   \includegraphics[width=1\textwidth]{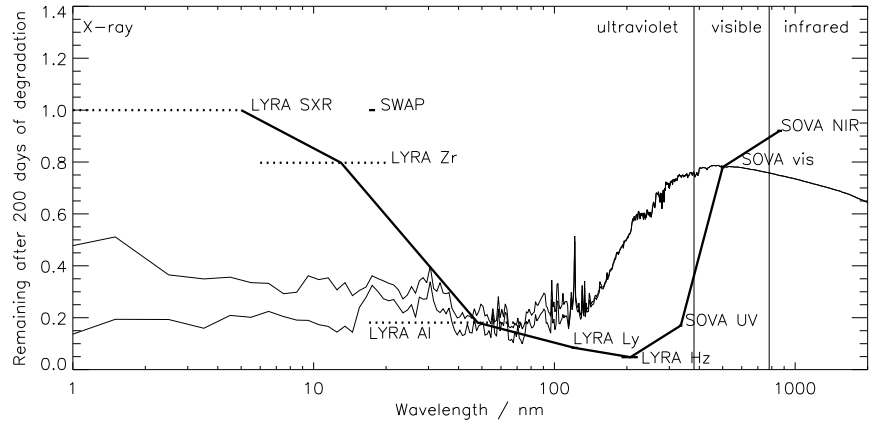}
   \caption{Normalized instrument response degradation as experienced by LYRA and SOVA after 200 days of open covers {\it vs.} spectral ranges of their individual channels.
            The two curves show two typical solar spectra on a log--log scale, one from a quiet-Sun, and one from maximum sunspot activity with Sun actively flaring, and
            are in arbitrary units.}
         \label{fig:lyra3}
 \end{figure}

The UV detectors of SOVA faced a degradation -- an immediate loss of $\approx$  70\,\% -- that appeared to be caused by radiation in space, independent of open-cover duration. 
Indeed high proton energies (from the SAA) induced secondary-particle generation when passing through the cover. In this case, the cover shielding is no longer effective.
This can be distinguished from the LYRA degradation and compared to the complementing LYRA channels in the UV range. 
Figure \ref{fig:lyra3} shows the degradation of the four LYRA unit-2 channels together with the three SOVA SPM-A channels, after 200 days of sunlight exposure.  
The solar spectra are plotted to demonstrate where the largest variability is, and what points or intervals the LYRA and SOVA channels actually correspond to.
The heaviest loss occurs in the UV around 200 nm; compared to this, the losses in the IR and SXR appear negligible. 

By connecting LYRA and SOVA data points, we suggest that there is probably a common mechanism responsible for the degradation of LYRA and SOVA,
most likely the contaminant deposited on the filters. 
The extreme degradation observed in the 20\,--\,500 nm range, mainly caused by some molecular contaminants,
implies further studies and strong requirements on mission preparation in order to avoid it in future long-term UV solar observations. 

LYRA appears to have avoided detector degradation by exploiting a different technology. 
Apart from this, the {\it window}  degradation -- obviously caused by UV-induced polymerization of contaminants on the filter surface -- has remained a problem since the times of SOVA. 
The experience with SOHO, launched in 1995, shows that it could have been avoided with an extensive cleanliness program.

\section {SDO/EVE Instrument Degradation}

The {\it Solar Dynamics Observatory} / {\it Extreme Ultraviolet Variability Experiment} (SDO/EVE: \opencite{eve}; \opencite{esp}) are solar EUV spectrometers, which show degradation of the EUV signal due to various mechanisms.

The SDO/EVE instrument comprises several channels using different technologies. 
The {\it EUV spectrophotometer} (ESP) is very similar to the SEM (cf. Section \ref{SEM}) using a transmission grating and photodiodes to provide zeroth-order and first-order measurements in the bands: 
0\,--\,7~nm (zeroth-order), 17\,--\,22~nm, 24\,--\,28~nm, 26\,--\,34~nm and 34\,--\,38~nm in first-order. Again an Al filter is used to restrict the bandpass incident on the grating. 
A separate Ti filter is used to further limit the bandpass seen by the zero-order detector. A Mg filter in front of the detector blocks second-order grating diffraction.

\begin{figure}[ht]
\centering
 \includegraphics[width=1\textwidth]{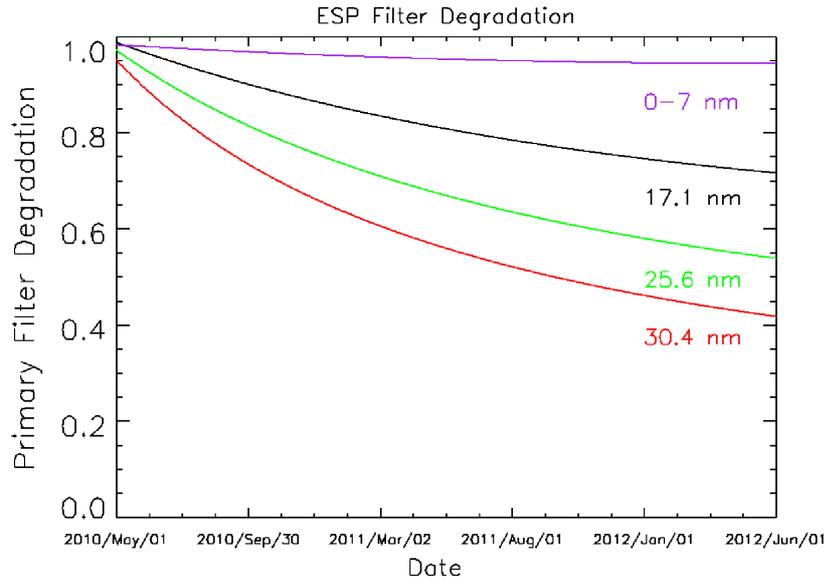}
 \caption{Degradation of the SDO-ESP channels. The ESP has three science filters in a filter wheel. 
By exposing the backup filters for only a very short amount of time, the degradation of the primary filter can be measured on-orbit.
 The calculated degradation is checked with rocket underflights.}
\label{fig:esp} 
\end{figure}

All channels of the ESP have shown degradation as shown in Figure~\ref{fig:esp}.
 This degradation has been shown to be due to front-filter contamination, as there are three Al filters on a filter wheel. 
The primary filter is used for most observations. The secondary filter is exposed for about five minutes a day to track the degradation of the primary filter,
 and a tertiary filter only exposed for five minutes a week tracks the degradation of the secondary filter. 
There is also a sounding rocket campaign that provides an independent determination of the degradation about once a year. 
Again, the degradation is consistent with a C layer forming on the front filter, and the thickness of C is appropriate for the degradation seen in all the wavelength channels.
  The 36~nm channel shows a different form of degradation due to a drastic reduction in the shunt resistance of the photodiode detector. 
It is not known what caused this failure, but the 36~nm channel has not returned useful data since launch.

\begin{figure}
 \centering
   \includegraphics[width=1\textwidth]{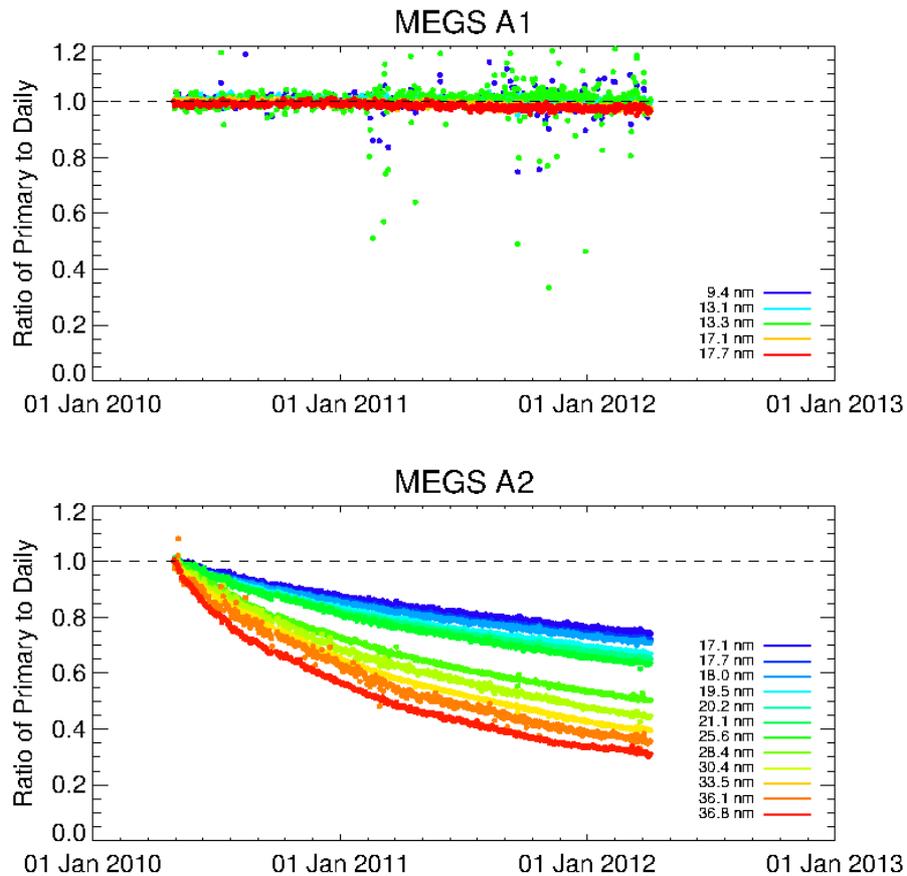}
  \caption{SDO/MEGS-A degradation.%
    Top: MEGS-A1 C--Zr--C filter shows insignificant degradation.
    Bottom: MEGS-A2 shows degradation very similar to that of ESP. Updated from \protect\opencite{Hock2012}.
      }
  \label{fig:MA1MA2}
\end{figure}

The {\it multiple EUV grating spectrograph} (MEGS) channels on EVE cover the 6--105~nm range with 0.1~nm resolution in three bands: 
MEGS A1, A2 and B. MEGS A1and A2 share a single grazing incidence mirror and grating. 
Two separate slits illuminate the grating each with separate thin-foil filter (also  of each type in a filter wheel.
 The A1 channel is optimised for the 5\,--\,18~nm range with a C--Zr--C filter. The A2 is optimised for 17\,--\,36~nm with a Al--Ge--C filter. 
The A2 channel shows degradation very similar to that experienced by ESP (Figure~\ref{fig:MA1MA2}, bottom), and a similar layer of C also explains the wavelength-dependent degradation.
 What is very interesting is that the A1 channel (only a few mm from A2) shows insignificant degradation (Figure~\ref{fig:MA1MA2}, top) even at 17~nm where there is overlap 
in the wavelength range between A1 and A2. There must therefore be something about the filter that causes contamination accumulation on the Al filter and not on the Zr.

\begin{figure}[ht]
\centering
   \includegraphics[width=1\textwidth]{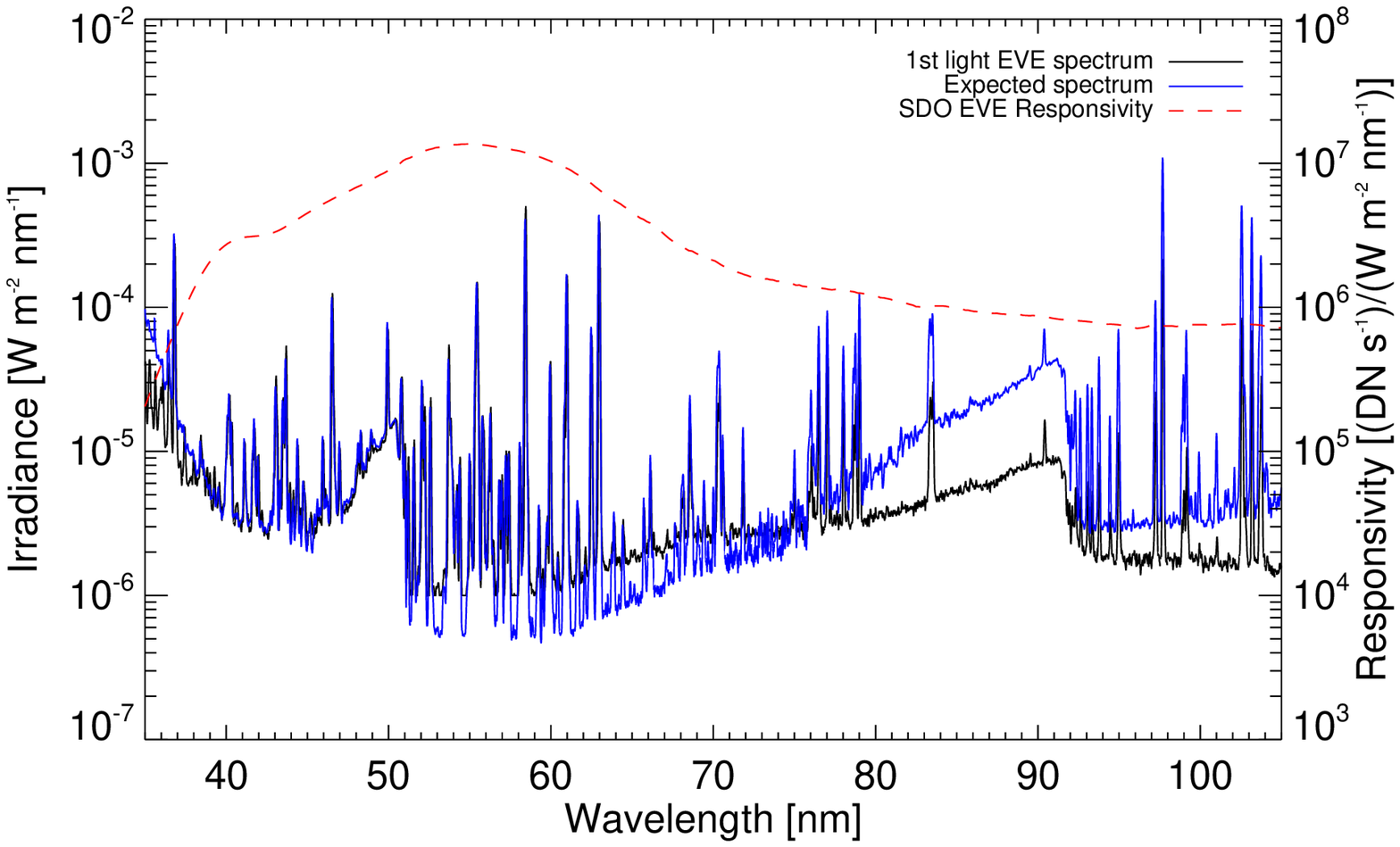}
 \caption{Degradation of the SDO/MEGS-B channel at first light. Updated from \protect\opencite{Hock2012}.}
\label{fig:MB} 
\end{figure} 

The MEGS-B channel is designed to operate in the 36\,--\,106 nm range. It is a cross-dispersed normal-incidence spectrometer, again with a CCD detector, nominally identical to the MEGS-A detector.
 MEGS-B does not use a filter-wheel filter for normal operation. MEGS-B showed dramatic degradation from the NIST-calibrated response at first light. 
Responsivity above  60 nm showed a steady drop and is about 90\,\% degraded at 105~nm (Figure ~\ref{fig:MB}). 
This ``first-light'' degradation could not be recovered by heating the detector to $+17\,^{\circ}\mathrm{C}$ (as hot as possible) for several days. 
The degradation continues to worsen with solar exposure, and flat-field images show ``burn-in'' of the brighter lines.
 It is thought that this degradation is due to back-side charging of the CCD. 
The initial charging was due to proton damage as SDO spent significantly longer in the geo-transfer orbit than planned,
 and this orbit dips into the inner proton belt twice a day, delivering a significant proton dose. 
The CCDs have a p-type implant to provide about a 7~nm  dead layer of Si between the SiO$_2$/Si interface and the charge collection region of the CCD.
 This was expected to provide enough isolation from back-side charging for the SDO mission life.
 However, once the potential due to surface charging exceeds that of the doping layer,
 degradation will be evident and will follow the penetration depth of Si (as seen in MEGS-B), and as the charging due to incident photons is dependent on the exposure time, 
this mechanism also explains the ``burn-in''. In order to maintain sensitivity of the MEGS-B channel it is only exposed for a short time each day, 
although campaign modes can be organized if continuous data is required. A similar effect is just beginning to be seen in the MEGS-A CCD, especially for the 30.4~nm line. 
However as the penetration depth of photons is deeper for the shorter wavelengths the effect is much less significant.

Finally MEGS-P is a lyman-$\alpha$ monitor. The zero-order from the first grating of MEGS-B is incident on an Acton~122XN interference filter to isolate the 121.6~nm lyman-$\alpha$ line. 
This is then measured by a photodiode. 
There is no noticeable degradation of this channel at all, which suggests that the MEGS-B first grating (exposed to the full solar spectrum) is not degrading either.

\section{Solar Instruments Onboard Picard}
\label{S-picard}
{\it Picard} is a scientific microsatellite (140 kilograms) that was launched on 15 June 2010. 
{\it Picard} is devoted to solar variability observation through imagery and radiometric measurements, with the aim of providing data for solar-physics investigation,
 and for the assessment of the influence of solar variability on the Earth's climate variability. 
The {\it PREcision MOnitoring Sensor} (PREMOS: \opencite{Schmutz2009}) and {\it SOlar Diameter Imager and Surface Mapper} (SODISM: \opencite{Meftah2010}),
the evolution of which is described in this article, are radiometers used to measure the solar irradiance and an imaging telescope to determine the solar diameter and asphericity, respectively.

\subsection{Operational Modes and Degradation of PREMOS}
PREMOS onboard the French satellite {\it Picard} comprises two experiments: 
one experiment is measuring the Total Solar Irradiance (TSI) with absolute radiometers, the other observing solar spectral irradiance at six different wavelengths with filter radiometers 
(given in Table \ref{premos-table}). We report below on the second experiment.
The optical and near-IR filters are identical to those in the {\it Picard}/SODISM instrument as well as the 215 nm filter,
 while the 210 nm filter was chosen to match the Herzberg channel implemented on the PROBA2/LYRA instrument. 
The PREMOS filter radiometer therefore covers an important part of the spectral range, which influences the chemical composition of the terrestrial atmosphere. 
The operational routines began on 6 September 2010 and PREMOS filter radiometers have provided continuous data since, even during eclipse season. 
Using a total of twelve channels divided into three instruments of four channels, 
a redundancy strategy has been established in order to estimate sensitivity loss due to exposure time to sunlight. 

One channel (Head A) is operated continuously (six measurements per minute with a integration time of 9.9 seconds for the normal mode), 
while its back-up channel (Head C) is exposed only once per day for three minutes. 
Finally, the Head B channel is a self-consistent system with duplicate channels; the first pair is exposed every fourth orbit for one minute,
 while the second pair is exposed once per week for about two minutes.

\begin{figure}
   \centering
  \includegraphics[width=1\textwidth,angle=90]{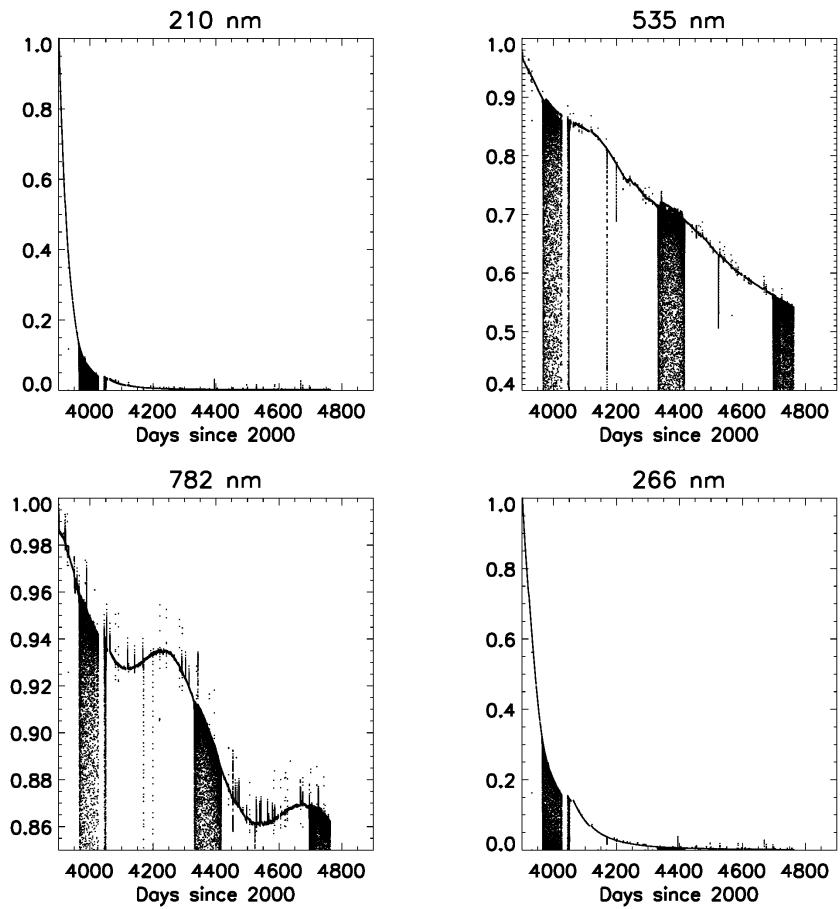}
   \caption{Normalized time series for all channels of Head A since the beginning of the PICARD mission.}
         \label{fig:premos1}
 \end{figure}

As displayed in Figure \ref{fig:premos1}, Head A has experienced a pronounced degradation since it has lost more than 99\,\% 
of the signal for the UV channels and about 35\,\% for the visible channel, while more than 86\,\% of the signal remains for the near-IR channel.
 We assume that this degradation is induced by the polymerization of contaminants on top of filters under the solar UV exposure. 
We are currently investigating why the degradation is not a decreasing function for the visible and near-IR channels. 
Head C has been exposed for only 40 hours since the beginning of the mission. 
For the UV channels (Head C), we estimate the loss of sensitivity to be about 10\,\% and 5\,\% respectively. It is much more difficult, 
however, to estimate the degradation of the visible and near IR channels.
The operational channels of Head B have been exposed to the Sun for approximately sixty hours, while the back-up channels have been exposed for less than four hours.
 We are currently using the channels of Head B to model the degradation for all UV channels.

\begin{table}
\caption{Wavelength characteristics of the PREMOS filter radiometers.}
\label{premos-table} 
\begin{tabular} {lcccc}                    
  \hline     
 &  channel 1 & channel 2 & channel 3 & channel 4 \\
\hline
HEAD A & 210 nm & 535 nm & 782 nm & 266 nm\\
HEAD B & 215 nm &  607 nm & 215 nm& 607 nm\\
HEAD C & 210 nm & 535 nm & 782 nm&  266 nm\\
\hline
\end{tabular}
\end{table}

\subsection{Aging of the Picard Payload Thermal Control: Impact on SODISM}
SODISM is an 11-centimeter Ritchey--Chretien imaging telescope developed by the French Centre National de la Recherche Scientifique (CNRS). SODISM measures the solar diameter and limb shape, and performs helioseismic observations to probe
 the solar interior. The solar diameter is measured at three wavelengths i.e. 535, 607 and 782 nm in the photospheric continuum. Images in the \ion{Ca}{ii} line (393 nm)
 are used to detect active regions near the solar limb, which may alter the diameter measurements. 
These images are also used to measure the solar differential rotation as well as to monitor space weather, together with images at 215 nm wavelength. 

Throughout the mission, thermal control ensures that each instrument or equipment unit is maintained at temperatures consistent with nominal operation. 
Most of the instruments only operate correctly if maintained at the right temperature and if temperature changes are within acceptable limits. 
Thermal control surfaces and optics of the payload are exposed to space environmental effects including contamination, atomic oxygen, UV radiation,
 and vacuum temperature cycling. 
The elements of SODISM that are regulated and not exposed to the Sun (\eg{} CCD, interference filters, mechanism, structure) remain stable with changes in temperature. 
In flight, the temperature of the SODISM's CCD (2028$\times$2048) is very stable, within $0.1\tc$. 

\begin{figure}
   \centering
   \vspace{0.5cm}
  \includegraphics[width=1\textwidth]{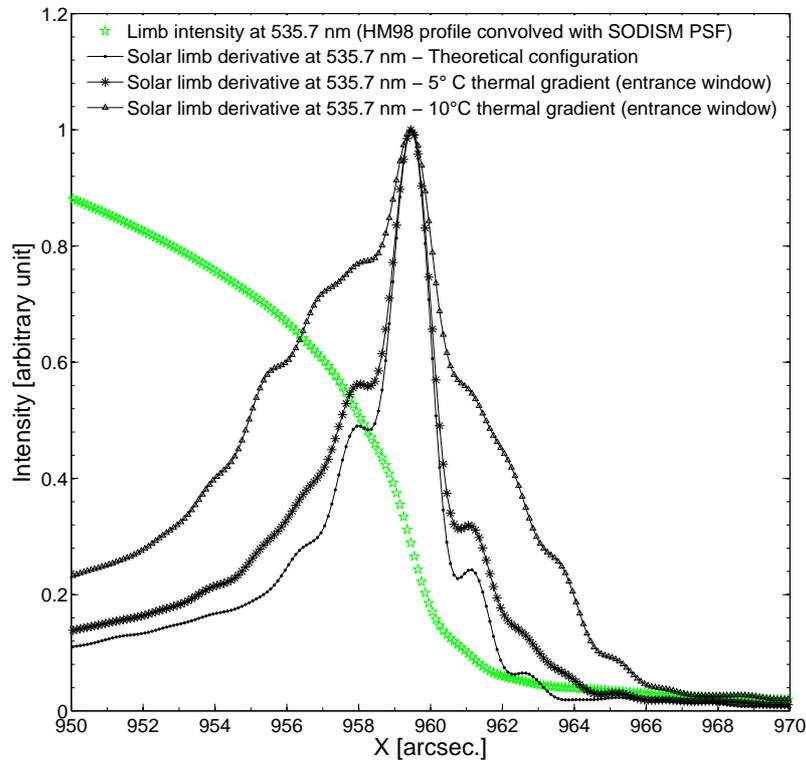}
   \caption{Thermal effect on the solar limb measurement at 535.7 nm.}
         \label{fig:sodism4}
 \end{figure}
 
Materials having low solar absorptance are often used for reflective surfaces designed to minimize heat absorption, 
but UV radiation degrades these materials by exponentially increasing the solar absorptivity of exposed surfaces. 
The presence of contamination on thermal control surfaces alters absorptance/emittance ratios and changes the thermal balance leading to an increase of the temperature 
of the payload. Contamination in optical instruments, on the other hand, reduces the signal throughput thus further reducing performance. 
The {\it Picard} payload thermal-control system includes several temperature-control techniques, such as reflective covers, coatings, insulation, and heat sinks. 
Aging of the covers, coatings, and insulation was observed and expected to be cumulative with time. The SODISM entrance window
 and the front of the instrument facing the Sun have a significant temperature increase. 
A general aging of the thermal-control system is observed as well.

The temperature of the SODISM front face varies greatly during an orbit and its temperature variation depends strongly on latitude and on the day of the year 
(variation and effect of incident fluxes).  This temperature evolution of the instrument front face principally impacts the main entrance window
and considerably degrades the measurement of the intensity profile's first derivative at the solar limb,
and consequently the measurement of the solar limb as illustrated in Figure \ref{fig:sodism4}.

\begin{figure}
   \centering
  \includegraphics[width=1\textwidth]{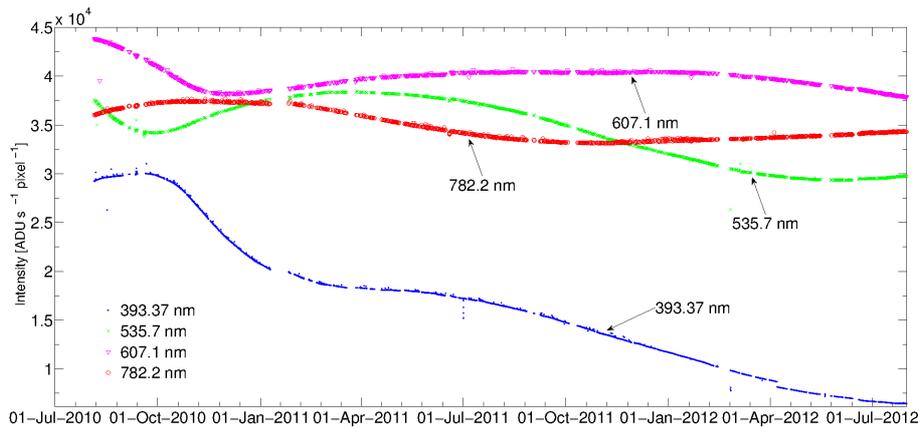}
   \caption{Temporal evolution of the SODISM CCD image intensity.}
         \label{fig:sodism3}
 \end{figure}

Despite the establishment of an active thermal control, there are also environmental effects on the SODISM instrument.
We observed that the CCD is strongly affected by SAA. The SODISM images, intensity (normalized to 1 AU) 
is shown in Figure \ref{fig:sodism3} and evolves over time with:
\begin{itemize}
\item Intensity oscillation at 535.7 nm, 607.1 nm and 782.2 nm,
\item Intensity oscillation and important degradation at 215 nm and 393.37 nm.
\end{itemize}
These effects might be  caused by the combination of contamination and degradation at the detector surface.
To outgas the accumulated contaminant on the CCD surface, a bake-out heater is installed on SODISM; this allowed periodic heating to +20$\tc$ 
during three days, but it was not effective. Another approach should be developed.

Thermal control, especially, for a payload suite, is crucial to mission success. 
For SODISM, a deterioration of the thermal-control system was observed in the long term (in particular in the front face of the telescope).
The measurements show a complex behavior with thermal and contamination effects as well. 
For the long-term evolution on SODISM measurements, we suspect a degradation of its CCD response, caused by contamination and energetic-particle issues,
and transmission filters at the entrance window of SODISM. 
Up to now, there is a good repeatability in measurements but most of the calibration requires thermal and/or optical corrections. 
Thermal coatings chosen for the {\it Picard} payload are adequate for maintaining temperatures in the acceptable range, but the use of radiators (white paints) 
facing the Sun is not the proper solution, a Sun shield with back-surface mirror should be used.
Moreover, metrology missions, such as {\it Picard}, require more dimensionally stable spacecraft structures.
Because there is a predominant effect between the latitude and the measurement of the solar radius, low-mass spacecraft in low orbits should be avoided. 

\section{Lessons Learned and Recommendations} 
\label{S-lessons-learnt}

The degradation of space instruments can be complex; their causes and mechanisms are, in many instances, difficult to understand, since they are often the result of the combination of several independent degradation processes. 
This fact is an especially important issue in establishing recommendations for best practices in developing and operating spaced-based solar instruments.
However, as demonstrated by the contributions of this article, 
the presence of contaminant species (organics and water) and exposure to radiation (both ionization and displacement-damage effects) are often the main reason for instrument degradation, 
and their impact is frequently underestimated. Contaminants can originate not only from spacecraft propellant, but also from outgassing or evaporation by all organic material used in the construction of these instruments. 
Further, once instruments are in space, the means to recover from degradation are very limited. 
For example, items that have collected contaminants while operating under cryogenic temperatures can be heated -- to an extent limited by electrical power available -- to desorb weakly bound molecules.
However, once molecules have settled by UV-polymerization, cleaning is very nearly impossible.
Although methods to recover from degradation have been established and successfully been used in the laboratory such as, \eg, UV-Ozone cleaning, such a method has not been implemented in a space mission.
Presently, there is no alternative to mitigation of contamination in space.

Our survey shows that three complementary strategies can dramatically minimize degradation and mitigate the effects of ongoing degradation:
\begin{itemize}
\item ensuring extreme cleanliness control during instrument development and launch, including careful material selection, minimization of organic material, and conditioning by bake-out,
\item monitoring the stability of the radiometric calibration using sophisticated methods,
\item identifying development needs for critical components (for example, imagers, photodetectors, optics, coating, electronics, {\it etc.}).
\end{itemize}

\subsection{Extreme Cleanliness Control} 

The cleanliness efforts for SOHO described in Section \ref{S-soho} were very successful; they were not excessive but neither were they completely perfect.
This became evident when the SOHO spacecraft was lost and then recovered after four months in 1998.
As the inter-calibration observations were resumed after the recovery of SOHO,
the degradation corrections for several instruments had to be completely remeasured because the temperature excursions during the phase of uncontrolled thermal environment released contaminants that had been residing for a long time on cold surfaces and resulted in a redistribution of contaminants and,
thus, accelerated instrumental degradation that had now to be taken into account.
So, while contaminants were still present onboard, the on-ground cleanliness activities reduced the potential for degradation considerably.

Our conclusion, therefore, is that stringent cleanliness efforts are an absolute prerequisite for calibration stability.
The main ingredients of a successful cleanliness program are the establishment of a cleanliness review board, inter-calibration working group, and instrument and spacecraft cleanliness control plans. 

The most important preventive measures for space instruments identified were:
\begin{enumerate}[i)]
\item Determination of the contamination sensitivity (also at spacecraft level) by modeling,
\item Design of the instrument to maximize cleanliness: 
\begin{itemize}
\item including design features such as purging concepts with large venting holes in the structural housing, the addition of vents and heaters close to 
the detector for bake-out ($>+35\tc$), door mechanisms, filter wheels with redundant filters to track front-filter contamination, contamination sensors (QCM), cold cup around the detector, solar-wind deflector plates, valves for depressurization during launch, {\it etc.}
\item selecting high-radiation tolerance and ultra-high vacuum quality materials with the lowest outgassing values available (the total mass loss, TML, and the collected volatile condensable materials,  CVCM, plus an additional parameter: the water vapor regained, WVR).
\end{itemize}
\item Stringent cleanliness procedures of all hardware:
\begin{itemize}
\item assembly in cleanroom (class ISO 5) with active charcoal filters,
\item use of oil-free vacuum systems during tests,
\item double bagging and continuous purging with pure and dry $\textrm{N}_2$ (grade 5.0) at instrument level up to the launch,
\end{itemize}
\item Extensive use of vacuum bake-outs at the component, sub-assembly, and final-assembly level:
\begin{itemize} 
\item baking at the highest temperature ($>+100\tc$) compatible with the material under clean gas purging,
\item bakeout durations determined by mass spectrometer and temperature-controlled QCM monitoring of cleanliness level,
\item gas-chromatography/mass spectrometry (GC/MS) analysis for acceptance/rejection of the components.
\end{itemize}
\end{enumerate} 
It should be added that the ground-support equipment that will be in direct contact with the flight hardware must be submitted to the same rules. As an example, cleanliness of vacuum chambers must be monitored by QCM or residual gas analysers. 

\subsection{Stability of the Instrument Radiometric Calibration} 
The careful and extensive radiometric calibration of the instruments prior to launch forms the basis of the success of spaced-based solar instruments. 
In most cases, opportunities for instrument-level tests and calibration are strongly limited by pre-launch scheduling constraints, but such testing is important enough that it warrants special consideration. 
Pre-flight calibration can be achieved with detectors and transfer radiation-source standards, both traceable to a primary standard source found in synchrotron-radiation facilities,
while the instruments themselves can be calibrated at the synchrotron facility or locally, at the instrument test facility, by transporting a transfer source standard to that facility. 
The latter option further reduces the possibility of contamination by exposing the instrument to an environment whose cleanliness cannot be sufficiently regulated. 

Once a spacecraft is in orbit, the stability of calibration can be monitored by carefully planned observations, but absolute calibration is often impossible.
Thus, a careful initial calibration and meticulous tracking of the evolution of instrumental calibration are both very important. 
Several different methods of calibration may be required to achieve this goal \cite{Schuehle2002}. 
Furthermore, qualified personnel and perhaps external expertise are often useful in the interpretation of data obtained, both on the ground and in-flight,
in order to accurately assess the evolution of the degradation of a space-based solar instrument.

\subsubsection{Onboard Calibration}

Onboard-calibration light sources have been essential to the success of many solar payloads and similar devices should always be included in the design of spaced-based solar instruments. 
Multiple calibration light sources (lamps or LEDs) may be carried onboard, and should be operated and exposed regularly to maintain an established calibration status.
It is worth noting that the value of calibration light sources is significantly reduced without pre-flight or pre-degradation reference data obtained during the on-ground calibration and in-flight commissioning phases of the mission.

However for EUV-VUV and X-ray instruments, for which calibration sources in the primary range of EUV instrumental sensitivity are not available,
the onboard visible-light flat-field that such onboard light sources provide can be used to monitor instrumental degradation if the relationship between visible and EUV degradation is known.
Although such lamps were used with great success in the in-flight calibration of EIT, these successes seem to be the exception rather than the rule.
Ideally, however, it should be possible to establish the relationships between visible and EUV degradation, as well as the potential for degradation of the light sources themselves,
before flight by irradiating engineering devices.
Additionally, we strongly recommend the use of blue or near-UV LEDs that have photon penetration depths similar to or lower than the EUV photons observed by the instrument,
and that every effort should be made to ensure complete and uniform detector illumination by these lamps on future EUV telescopes.

\subsubsection{Calibration Updates and Inter-Calibration}
Alternatively, it is possible to track instrumental calibration by inter-calibration using observations from occasional rocket underflights using similar instruments that can be carefully calibrated on the ground both before and after the flight.
Another option for establishing absolute calibration using in-flight observations is the use of invariant sources -- assuming they are accessible by the instrument -- such as observations of celestial standard sources, or of Sun-center during quiet periods,
or by inter-calibration of identical variable sources using different instruments with similar, corresponding wavelength sensitivity.
In the case of visible-light imaging instruments, observations of the background star field can also be used to monitor long-term degradation,
as has been done very successfully in the case of STEREO/SECCHI's HI-1 imager.
Note, however, that for inter-calibration measurements, off-point and roll maneuvers of the spacecraft platform may be required to correct for the effects of spatial and spectral dependent degradation patterns (i.e. flat field and stray light).

\subsubsection{Redundancy Concept}
Redundancy can be implemented at either component or instrument levels (such as LYRA and PREMOS).
Past experience shows that redundancy is useful in recovery from degradation and damage from different causes.
For example, the degradation of thin-film filters typically occurs during launch,
but impact by debris or micrometeoroids (see, for example, EIT) can cause irreparable damage if redundant parts are not available.
For radiometers, the use of several spare units -- the number may depend on the projected lifetime of the mission -- with separate door covers is strongly recommended.
In addition to protecting against potential single-unit failures, observations by less-exposed units with a low duty-cycle -- for example, a few minutes per week -- can provide valuable information on the long-term evolution of the instrument.

\subsection{Identification of Development Needs} 

Specific design and technological development  is particularly important for UV instruments.
For the SOHO UV instruments, the optical systems were quite stable during flight \cite{Schuehle2003}, but the detectors remained a source of instability \cite{thompson99}.
This was partly due to the temperatures of CCD detectors and partly due to the effect of irradiation on the Si devices, while detectors with multichannel-plate intensifiers suffered from gain depletion during exposures. 

This detector-degradation issue is frequently foreseen before launch, but both its importance and severity often underestimated.
A list of all proven technologies and their degradation levels is beyond the scope of this article,
but one especially important recommendation concerns the use of back-illuminated detectors (and affects both CCDs and CMOS APS detectors equally).
The commonly used detector interface (Si/SiO$_2$) is very sensitive to radiation damage (both by protons and UV photons) in space,
which leads to a decline in detector sensitivity over time. However, proper surface passivation of the backside of existing detectors can reduce the impact of radiation exposure,
and intensive pre-flight characterization can help mitigate the damage that cannot be controlled.
It would be worthwhile to explore the use of both alternative oxides with greater radiation tolerance (\eg{} Al$_2$O$_3$) and non-oxide passivation layers. 

In most cases, CCDs should be kept at the lowest possible operating temperature in order to reduce dark current,
the effects of radiation damage, and the appearance of \textit{hot, warm}, and \textit{flipping} pixels.
Our analysis suggests that operating temperatures should be lower than $-60\tc$ for non-inverting mode operation (NIMO) or $-40\tc$ for asymmetric inverting mode operation (AIMO) CCDs.

Finally, even while research and development in space technology is widely acknowledged as essential for designing future long-lifetime space missions,
we recommend intensified efforts to develop advanced photon radiation detection systems, in particular: 

\subsubsection{Next-Generation CMOS-APS}
We expect that many future instruments will make use of highly efficient CMOS-APS detectors similar to the one used by SWAP.
In fact, the {\it Extreme Ultraviolet Imager} (EUI), the {\it Heliospheric Imager} (SoloHI),
and the {\it Polarimetric and Helioseismic Imager} (PHI) onboard the {\it Solar Orbiter} mission are all expected to incorporate
the next-generation of CMOS-APS detectors with significantly improved characteristics (see \opencite{BenMoussa2013} for the detector prototypes development of EUI). 

\subsubsection{Thoroughly Tested UV Filters}
There is a great need for optical elements of all kinds (filters, grazing reflectors, and mirrors) with both improved radiation tolerance and spectral purity.
This need is demonstrated by the rapid degradation of the UV filters on both PROBA2 and {\it Picard}.
For successful future missions, both modeling and complete test campaigns for UV and visible filters (including radiation and contamination simulations tests) are basic requirements.

\subsubsection{Radiation-Hardened UV-Sensitive Materials}
Radiation hardness against UV photons or protons is also a primary concern for upcoming long solar missions residing several years in space.
There are promising alternatives to the commonly used silicon-based imagers and photodetectors based on wide band-gap materials such as the diamond detectors used in LYRA.
Details of these next-generation detectors are discussed by \inlinecite{BenMoussa2009b}.
A proof-of-concept AlGaN imager ($256 \times 256$~pixels), sensitive only to UV and operating at room temperature, has been recently demonstrated by the {\it Blind to Optical Light Detectors} project (BOLD: \opencite{Malinowski2011}).

\subsubsection{Onboard Data Processing}
Given the issues that remain in providing high data-flow for nearly all space-based instruments, and, in particular,
issues with the optimization of data-flow for spacecraft in low-telemetry orbits, future systems must be capable of high-performance onboard computing, which, in turn,
requires high-performance, radiation-hardened field-programmable gate arrays (FPGAs) that can perform automated onboard calibration.
For some missions it will be necessary to update detector-calibration maps and perform onboard image correction such as high-quality cosmic-ray removal in order to prevent unrecoverable distortion caused by low-quality image compression or a very poor lossless compression ratio. 

\section{Conclusion} 
\label{S-Conclusion} 
The workshop that took place in the Solar Terrestrial Centre of Excellence (STCE) in Brussels, Belgium on 3 May 2012,
provided an excellent starting point for dialog between experts and facilitated the exchange of much experience gained during space-based solar missions.
The outcomes of this meeting and discussion, together with the written contributions of the different mission teams, have sparked this article focusing on the major lessons learned about in-orbit degradation of solar instruments.

Although this article addresses scientists and, perhaps more specifically, engineers involved in spaced-based solar instrument development,
all stakeholders of any project should be deeply involved in the assessment and monitoring of any degradation, and, since the consequences of degradation can be quite severe,
should take the issue extremely seriously. 

There are several approaches to assessing and monitoring the degradation of spaced-based solar instruments that give good results,
many of which we have discussed above. A prime conclusion of this work is that there is no single best method, but rather
that a combination of methods must be critically selected, taking into account the applicability of the methods given both the mission targets and instrumental design itself.
It is therefore important to continue to share regular and open information about what is working and what is not, in order to learn from the community's shared experiences.

In particular, identifying the lessons learned from past projects is of special value to both the community and instrument teams themselves.
Unfortunately, project teams often move quickly from project to project, and identifying the lessons learned rarely seems to be a priority.
With this article, we hope to address this problem directly.
We have identified the lessons learned by a broad range of instruments and missions that comprise a vast range of solar-physics objectives and span nearly two decades of experience. We hope that these lessons can be ingested by new instrument development teams and, in turn,
can prevent both current and future missions from repeating past mistakes. It is the motivation of each individual (scientist and engineers) to learn,
share, and change what makes the lesson learned successful. Prevention is far better and much cheaper than cure.

\begin{acks}
The authors acknowledge the support from the Belgian Federal Science Policy Office (BELSPO)
through the Solar Terrestrial Centre of Excellence (STCE) program.
\end{acks}

\end{article} 
\end{document}